\documentclass[]{article}

\usepackage{graphicx}
\usepackage{cancel}
\usepackage{authblk}
\usepackage{amsmath}
\usepackage{amsthm}
\usepackage{amssymb}

\usepackage[sort&compress,numbers]{natbib}
\usepackage[colorlinks=true,linkcolor=black, citecolor=blue, urlcolor=blue]{hyperref}

\begin{document}


\title{Displacement general solutions in strain gradient elasticity: review and analysis} 

\author[1,2]{Y. Solyaev}
\author[2,3]{E. Hamouda}
\author[4]{S. Sherbakov}

\affil[1]{Institute of Applied Mechanics of Russian Academy of Sciences, Moscow, Russia}
\affil[2]{Moscow Aviation Institute, Moscow, Russia}
\affil[3]{Mansoura University, Mansoura, Egypt}
\affil[4]{Belarussian State University, Minsk, Belarus}

\setcounter{Maxaffil}{0}
\renewcommand\Affilfont{\itshape\small}

\date{\today}

\maketitle

\begin{abstract}
In this work, we provide an overview of general solutions for displacement fields in static problems of isotropic strain gradient elasticity (SGE). We not only review existing solutions but also derive new representations, showing that all classical elasticity solutions -- including those of Boussinesq-Galerkin, Papkovich-Neuber, Naghdi, Lam\'e, Love and Boussinesq -- can be simply generalized to SGE framework. 
In general, it is shown that SGE enables the use of any classical general solution representation combined with a Helmholtz decomposition for the gradient part of the displacement field. Consistency is also established between the presented Papkovich-Neuber representation and the general solutions of SGE proposed previously by Mindlin (1964), Lurie et al. (2006) and Charalambopoulos et al. (2020). Furthermore, we establish the relationships between the stress functions of different general solutions and show their completeness. 
\end{abstract}

\section{Introduction}
\label{int}

The representations of displacement general solutions for equilibrium equations of linear elastic bodies developed by Boussinesq \cite{boussinesq1885application}, Galerkin \cite{galerkin1930investigation}, Papkovich \cite{papkovich1932representation} and Neuber \cite{neuber1934neuer}, Ter-Mkrtychan \cite{ter1947general} and Naghdi and Hsu \cite{naghdi1961representation}, etc., have enabled the derivation of the broad range of analytical solutions for important applied problems (see, e.g. \cite{lur1964three,lurie2010theory, barber2004elasticity}). The development of operator formalism for general solutions derivation along with the discussion on their completeness, non-uniqueness, and interrelations between different forms of general solutions and different stress functions was an important topic of the research in classical elasticity \cite{mindlin1936note, lur1937theory, eubanks1956completeness, gurtin1962helmholtz, wang1991transformation, wang1995completeness, wang2008recent}.

In the present study, we consider the generalized formulation of static  theory for isotropic linear elastic bodies developed by Toupin \cite{toupin1964theories} and  Mindlin \cite{Mindlin1964} (see also \cite{germain1973method, green1964multipolar, dell2017higher}). This theory assumes that the strain energy density of material depends on strain and on the strain gradients and it is called the strain gradient elasticity (SGE) or the second gradient elasticity or the elasticity for solids of grade two \cite{toupin1964theories, dell2017higher}. There are differences in the formulation of the constitutive relations of isotropic SGE, which can be written in terms of the second gradients of displacements (Mindlin Form I), gradients of strains (Mindlin Form II), or in terms of the rotation gradient and the fully symmetric part of the second gradients of displacements (Mindlin Form III) \cite{Mindlin1964}. Nevertheless, all these variants lead to the same form of the equilibrium equations, which are of a higher order and contain two additional length scale parameters \cite{Mindlin1964}: 
\begin{equation}
\alpha \left( 1-l_{1}^{2}\nabla ^{2}\right) \nabla \nabla \cdot \mathbf{u}%
-\left( 1-l_{2}^{2}\nabla ^{2}\right) \nabla \times \nabla \times \mathbf{u}%
=-\frac{\mathbf{b}}{\mu } 
\label{ee}
\end{equation}
where $\mathbf{u}(\textbf r)$ is the displacement field, $\mathbf{b}(\textbf r)$ is the body force density, $\textbf r$ is the position vector, $\mu $ is the Lam\'{e} parameter (shear modulus),
$\alpha =2\left( 1-\upsilon \right) /\left( 1-2\upsilon \right)$, $\upsilon $ is Poisson's ratio; $l_{1},l_{2}$ are the length scale parameters of isotropic elastic material. Definitions of $l_{1}$ and $l_{2}$ in terms of classical and gradient material constants depend on the constitutive equations of the considered gradient theory \cite{Mindlin1964, dell2009generalized}. When $l_{1}=l_{2}=0$ the formulation reduces to classical elasticity theory.

In this introductory section, we provide a brief historical overview of the development and applications of general solutions for SGE equilibrium equations \eqref{ee}. A detailed mathematical formulation of these solutions will be presented and analyzed in Sections 3 and 4.

Mindlin has developed the first variant of general solution representation for equations \eqref{ee} and applied it for the Kelvin problem solution within SGE in 1964 \cite{Mindlin1964}. Mindlin representation was exactly the generalization of solution that was developed previously within the couple stress theory, which equilibrium equations are the particular case of \eqref{ee}, when $l_1 = 0$ \cite{mindlin1962effects}.
The Mindlin general solution and its particular variants for centro- and axial-symmetric problems were used later in the analysis of stress concentration around cylindrical and spherical cavities and inclusions \cite{hazen1968stress, weitsman1966strain, eshel1970effects, eshel1973some}. Its particular variant for the simplified theory with $l_1=l_2$ was used recently for the Kirsch problem solution \cite{khakalo2017gradient}. The strong size effect and the reduction of stress concentration around the small-size holes and cavities have been established \cite{hazen1968stress, weitsman1966strain, eshel1970effects, eshel1973some,khakalo2017gradient}. The size of the holes is compared in these solutions to the length scale parameters of material.

Later, another form of general solution was proposed in the works of Lurie and co-authors \cite{lurie2006interphase,lurie2009advanced,volkov2010eshelby,lurie2011eshelby}. This representation was obtained independently and, unlike Mindlin's solution, it explicitly separated the classical and gradient parts of the displacement field, the sum of which determines the total displacements in the elastic medium. The solution by Lurie et al. was initially proposed for a simplified variant of constitutive relations of SGE and it was later generalized to various formulations of SGE, including the case of the general form of equilibrium equations \eqref{ee} \cite{lurie2023general}. This solutions were used in the various micromechanics problems for evaluation of the effective properties of composites materials accounting for size effect and redistribution of stress field around nano-sized inclusions \cite{lurie2011eshelby, lurie2016exact, lurie2016multiscale, solyaev2019three}.

Charalambopoulos and Polyzos have considered the simplified variant of Mindlin theory with $l_1=l_2$ in \eqref{ee}. For this case, they have shown that any kind of SGE general solutions can be presented as the sum of a classical displacement field, and an additional field that simply obeys the Helmholtz equation \cite{charalambopoulos2015plane, charalambopoulos2022representing}. In Section 3.8 of the present paper, this result will be generalized for the general formulation of SGE with $l_1\neq l_2$ in \eqref{ee}.  Charalambopoulos et al. have also developed the simplified form of Mindlin general solution \cite{charalambopoulos2020plane}  and applied this representation for the analysis of the plane strain tension and bending problems for rectangular domains \cite{charalambopoulos2015plane, charalambopoulos2022representing, charalambopoulos2020plane}.

Note that all the mentioned general solutions \cite{Mindlin1964,lurie2009advanced, charalambopoulos2020plane} can be considered as generalizations of the classical Papkovich-Neuber representation, although they include additional potentials and the displacement field depends on these potentials in a more complex way. However, in the present work, we will denote the generalized Papkovich-Neuber solution within SGE the form that was proposed in works \cite{solyaev2022elastic, solyaev2024complete}. This representation turns out to be the closest to the classical one -- it explicitly contains the classical Papkovich-Neuber representation, as well as a minimal number of additional potentials responsible for the gradient part of the displacement field. Furthermore, this representation is obtained for the general formulation of SGE, and its completeness was demonstrated. The proof of completeness, as in the classical case, is based on using the general solution in the Boussinesq-Galerkin form, which for equation \eqref{ee} was also presented in work \cite{solyaev2024complete}.

The generalized Papkovich-Neuber solution \cite{solyaev2022elastic, solyaev2024complete} has been applied within the framework of SGE to several problems. These include the asymptotic analysis of a plane-strain problem of the wedge under a concentrated load \cite{solyaev2022elastic} and crack problems, including a generalized higher-order Williams' solution \cite{solyaev2023application,solyaev2024higher}. The full-field three-dimensional solutions have also been derived for Kelvin problem \cite{solyaev2024complete} and for the problem of a sphere subjected to an equatorial line load \cite{dell2024deformation}. The regularization of displacement, strain and stress fields in these problems has been shown that is in agreement with the other known results obtained by using another analytical and numerical methods in SGE (see, e.g. \cite{gourgiotis2009plane, askes2011gradient, papanicolopulos2010numerical, gourgiotis2018concentrated, lazar2018mindlin, vasiliev2021flamant, chirkov2024plane}).

The Lam\'e's strain potential method was developed in gradient elasticity by Papargyri-Beskou and Tsinopoulos \cite{papargyri2015lame}. This method is limited to the rotation-free problems and it was applied for plane problems of tension of rectangular domains and axisymmetric deformations of cylindrical domains under prescribed kinematic and static boundary conditions.  Authors considered the simplified SGE assuming $l_1=l_2$ in \eqref{ee}, though the representation can be simply extended to the general theory. Below we will show that this solution representation, along with other commonly used specific general solutions for axisymmetric problems (Boussinesq and Love), can be readily derived from the other more general representations, just as in classical elasticity theory.

The rest of the paper is organized as follows. In Section 2, we provide preliminary definitions and representations that are well known but are included explicitly to make the paper self-contained. In Section 3, we present ten different representations of the general solutions in SGE. Several of these are new and have not previously been considered in SGE (the generalized Naghdi, Boussinesq, and Love solutions). For these new representations, we provide their derivation; for the others, we give only the expression for the displacement field in terms of potentials and the governing equations for potentials, with references to their original derivations. In Section 4, we present the interrelations between the potentials (also referred to as "stress functions" \cite{gurtin1972}) of different general solutions. The most important and novel result is the established relationship between the Mindlin general solution \cite{Mindlin1964} and the generalized Papkovich-Neuber solution \cite{solyaev2022elastic,solyaev2024complete}. This relationship provides an automatic correspondence between all known general solution forms within SGE, including those by Lurie et al. \cite{lurie2006interphase} and Charalambopoulos et al. \cite{charalambopoulos2022representing}. Finally, in Section 5, we provide a simple proof of completeness for all the considered general solutions. 

\section{Preliminaries}

When constructing solutions for gradient theories of grade two (in any area of continuum mechanics and physics \cite{solyaev2024complete, lazar2014gradient,dimitrova2020quadrupolarizability}), one frequently encounters equations of the following form:
\begin{equation}
\left( 1-l^{2}\nabla ^{2}\right) \nabla ^{2}\mathcal{F}=\mathcal{B},
\label{P1}
\end{equation}%
where $\mathcal{F}$ and $\mathcal{B}$ are the scalar (or vector) fields ($\mathcal{F}$ is the field variable to be determined, while $\mathcal{B}$ is the prescribed right hand side); and the length scale parameter is real ($l\in \mathbb{R}$)  so that the modified Helmholtz equation is represented by the operator $\left( 1-l^{2}\nabla ^{2}\right) $.

General solution of equation \eqref{P1} can be presented in the following form:
\begin{equation}
\begin{aligned}
\label{bf}
\mathcal{F}=\mathcal{F}_{c}+\mathcal{F}_{g},\qquad
\nabla ^{2}\mathcal{F}_{c}=\mathcal{B}, \qquad \left( 1-l^{2}\nabla
^{2}\right) \mathcal{F}_{g}=l^{2}\mathcal{B}
\end{aligned}
\end{equation}
in which $\mathcal{F}_{c}$ and $\mathcal{F}_{g}$ are the general solutions of
the Poisson equation and inhomogeneous modified Helmholtz equation,
respectively.

Representation \eqref{bf} is useful since the general solutions and many kinds of particular solutions of obtained classical second-order equations are well known \cite{morse1954methods, zielinski1985generalized}.
Representation \eqref{bf} is also complete in the sense that for any solution $\mathcal{F}$ of equation \eqref{P1} there exist the corresponding solutions $\mathcal{F}_{c}$ and $\mathcal{F}_{g}$ that can be found in the following form: 
\begin{equation}
\begin{aligned}
\label{bfc}
\mathcal{F}_{c}=\left( 1-l^{2}\nabla ^{2}\right) \mathcal{F},\qquad
\mathcal{F}_{g}=l^{2}\nabla ^{2}\mathcal{F}
\end{aligned}
\end{equation}

The proof of representations \eqref{bf} and \eqref{bfc} is straightforward and can be found, e.g., in \cite{solyaev2024complete, lazar2014gradient}.

In the following derivations we will involve the Green's functions for determination of particular solutions of the Poisson equation and  inhomogeneous modified Helmholtz equation that are given by the following convolutions (in three-dimensional space):

\begin{equation}
\mathcal{N}\left( \mathcal{B} \right) =-\frac{1}{4\pi }\int_{D}%
\frac{\mathcal{B}\left( \mathbf{\pmb\xi }\right) }{\left\vert \mathbf{r}-\pmb\xi%
\right\vert }\mathrm{d}\pmb\xi
\label{gle}
\end{equation}

\begin{equation}
\mathcal{H}\left( \mathcal{B} \right) =\frac{1}{4\pi }\int_{D}%
\frac{\mathrm{e}^{-\left\vert \mathbf{r}-\pmb\xi\right\vert
/l}\mathcal{B}(\pmb\xi) }{\left\vert \mathbf{r}-\pmb\xi
\right\vert }\mathrm{d}\pmb\xi
\label{ghe}
\end{equation}

Finally, we will also refer to Helmholtz theorem that states that every sufficiently smooth vector field $
\mathbf{u}\left( \mathbf{r}\right) $ (with appropriate behaviour at infinity \cite{gurtin1962helmholtz,petrascheck2015helmholtz}) can be written as follows:%
\begin{equation}
\mathbf{u}=\nabla \varphi +\nabla \times \textbf A
\label{ht}
\end{equation}%
where $\varphi$ is the scalar potential, $\textbf A$ is the vector
potential that is assumed to be solenoidal ($\nabla \cdot \textbf A=0$) without loss of generality, and it is valid that  
\begin{equation}
\nabla^2 \varphi = \nabla\cdot\mathbf{u},\qquad
\varphi = \mathcal N (\nabla\cdot\mathbf{u}), \qquad
\nabla^2 \textbf A = -\nabla\times\mathbf{u}
\label{P3}
\end{equation}
in which the integral convolution operator $\mathcal N$ is defined according to \eqref{gle}.

In classical elasticity, the fields $\varphi$ and $\textbf A$ should satisfy the Poisson equations. This result follows from classical equilibrium equations that can be obtained from \eqref{ee} assuming $l_1=l_2=0$. Thus, using \eqref{ht}, \eqref{P3} and similar Helmholtz decomposition for the body force ($\textbf b = \nabla \varphi_b +\nabla \times \textbf A_b$) in classical elasticity we have \cite{lurie2010theory}:
\begin{equation}
\alpha \nabla \nabla \cdot \mathbf{u}
- \nabla \times \nabla \times \mathbf{u}
=-\frac{\mathbf{b}}{\mu }
\quad\implies\quad
\alpha \mu \nabla^2 \varphi =-\varphi_b, \quad
\mu \nabla^2 \textbf A =-\textbf A_b
\label{eec}
\end{equation}

The direct application of the Helmholtz theorem to static problems of classical elasticity is limited to incompressible materials \cite{gurtin1972,lurie2010theory}. This limitation arises because, according to \eqref{P3} and \eqref{eec}, the displacement field must be divergence-free when $\mathbf{b}= 0$ (or at least when $\varphi_b=0$).
The Helmholtz theorem's utility increases in the context of SGE. Here, it offers a direct representation of the general solution for the gradient part of the displacement field. Charalambopoulos et al. \cite{charalambopoulos2022representing} first highlighted this application for a simplified SGE (with $l_1=l_2$). Section 3.8 of this paper will address the formulation for the general theory. An example of direct application of decomposition of the type \eqref{ht} within plane problems of SGE was considered in Ref. \cite{placidi2017semi}.

\textbf{Notations}.  
In the following sections we employ direct notation, with \textbf{boldface} for vectors and \textit{italics} for scalars. Subscripts "$c$" and "$g$" denote the classical and gradient parts, respectively, of field variables (e.g., displacement, stress functions). An asterisk superscript (*) indicates particular solutions and related fields.

\section{General solution representations in SGE}

\subsection{Mindlin solution}

Mindlin \cite{Mindlin1964} derived the following form of general solution for equation \eqref{ee}:
\begin{equation}
\mathbf{u}=\mathbf{B}-l_{2}^{2}\nabla \nabla \cdot \mathbf{B}-\tfrac{1}{2}%
\left( \kappa -l_{1}^{2}\nabla ^{2}\right) \nabla \left( \mathbf{r}\cdot
\left( 1-l_{2}^{2}\nabla ^{2}\right) \mathbf{B}+B_0\right) ,
\label{mgs}
\end{equation}
where $\kappa = \frac{\lambda+\mu}{\lambda+2\mu} = \tfrac{1}{2(1-\nu)}$ and the stress functions $\mathbf{B}$ and $B_0$ have to
satisfy the equations
\begin{equation}
\begin{aligned}
\left( 1-l_{2}^{2}\nabla ^{2}\right) \nabla ^{2}\mathbf{B} &=-\frac{\mathbf{%
b}}{\mu }, \\
\left( 1-l_{1}^{2}\nabla ^{2}\right) \nabla ^{2}B_0 &=%
\frac{1}{\mu }\left( \mathbf{r}\cdot \left( 1-l_{1}^{2}\nabla ^{2}\right)
\mathbf{b}-4l_{1}^{2}\nabla \cdot \mathbf{b}\right) 
\label{mgse}
\end{aligned}
\end{equation}

When $l_1=l_2=0$ representation \eqref{mgs} reduces to classical Papkovich-Neuber solution. However, within SGE this representation proves to be quite complex. First, the governing equations for the potentials \eqref{mgse} are of the fourth order. While this issue is easily resolvable, and the representation can be simplified by using \eqref{P1} and \eqref{bf}, a more significant difficulty lies in expressing the displacement solution through high-order derivatives of the potentials. In \eqref	{mgs}, it is necessary to compute derivatives up to the fifth order, which is rather inconvenient from a computational standpoint. Consequently, subsequent work in this field has aimed at simplifying representation \eqref{mgs}.

\subsection{Lurie-Belov-Volkov-Bogorodskiy (LBVB) solution}

The possibility of representing the general solution to an equation \eqref{ee} as a sum of classical and gradient parts appears to have been first established in Refs. \cite{lurie2006interphase, lurie2009advanced}. This representation was initially proposed for a simplified gradient theory with a single length scale parameter. It was later generalized to equilibrium equations of Mindlin Form II \eqref{ee} as follows \cite{solyaev2019three, lurie2023general}:
\begin{equation}
\label{lgs}
\begin{aligned}
	\textbf{u} &= \textbf{u}_c + \textbf{u}_g,\\
	\textbf{u}_c &= \textbf F_0
			- \tfrac{\kappa}{2} \nabla (\textbf{r} \cdot \textbf F_0 + f),\\
	\textbf{u}_g &=  \textbf F_2 
			- \nabla\nabla\cdot (l_1^2\textbf F_1 + l_2^2\textbf F_2)
\end{aligned}
\end{equation}
where $\textbf{u}_c$ is the classical part of the solution, which is defined through the  classical Papkovich-Neuber harmonic vector potential $\textbf F_0$ and scalar potential $f$;  $\textbf{u}_g$ is the gradient part of the solution that is defined through the two additional vector potentials $\textbf F_1$, $\textbf F_2$. 

In absence of body forces, classical potentials should satisfy the Laplace equation:
\begin{equation}
\label{lgsce}
\begin{aligned}
	\nabla^2 \textbf F_0 = 0, \qquad \nabla^2 f = 0
\end{aligned}
\end{equation}
while the gradient potentials should obey the modified Helmholtz equations:
\begin{equation}
\label{lgsge}
\begin{aligned}
	\textbf F_1-l_1^2 \nabla^2\textbf F_1 = 0,\qquad
	\textbf F_2-l_2^2 \nabla^2\textbf F_2 = 0
\end{aligned}
\end{equation}

A representation for the case of nonzero body forces can be found in a recent publication \cite{lurie2023general}. Since it employs different notation for the length scale parameters, we omit it here. An additional reason for this omission is that the representation to be introduced in Section 3.5 is very similar and is already formulated for nonzero body forces. For comparing these two solutions, the representation presented in Eq. \eqref{lgs}-\eqref{lgsge} will be sufficient.

We can also note the redundancy in the number of independent functions used to represent the gradient part of the displacement field in \eqref{lgs}-\eqref{lgsge}. Clearly, the vector function $\textbf F_1$ can be replaced by a scalar, since only its divergence is involved in the representation \eqref{lgs}.
Thus, this representation involves four functions to determine the classical part of the displacement field ($\textbf F_0$, $f$) and an additional four independent quantities ($\nabla\cdot\textbf F_1$, $\textbf F_2$) for its gradient part.

As is known, for the classical Papkovich-Neuber representation (and consequently, for the classical part of the displacement field $\textbf u_c$ in \eqref{lgs}), the use of four independent functions is necessary only for the case $\nu = 0.25$ \cite{lurie2010theory}. No such constraint applies to the gradient part of the displacement field $\textbf u_g$. Therefore, to represent its three components, it is sufficient to use only three independent functions.
This fact prompted the subsequent modification of the representation \eqref{lgs}, which eliminated the redundant number of potentials for the gradient part of the displacement field (see Section 3.5).

\subsection{Charalambopoulos-Gortsas-Polyzos (CGP) solution}

In the paper \cite{charalambopoulos2020plane}, the authors proposed the simplification of Mindlin's general solution for SGE with single length scale parameter ($l_1=l_2=l$). It was shown that the displacement representation \eqref{mgs} can be replaced by the following:
\begin{equation}
\mathbf{u}=\mathbf{B}-\tfrac{\kappa}{2}%
\nabla \left( \mathbf{r}\cdot
\left( 1-l^2\nabla ^{2}\right) \mathbf{B}+B_0\right) ,
\label{cgps}
\end{equation}%
where the stress functions $\mathbf{B}$ and $B_0$ are the same as those in the Mindlin solution \eqref{mgs} and satisfy governing equations \eqref{mgse} under condition $l_1=l_2=l$.
 
This representation of the general solution \eqref{cgps} involves third-order derivatives of the potentials, therefore it is simpler than the Mindlin solution \eqref{mgs}. Furthermore, considering representation \eqref{bf}, \eqref{bfc} for \eqref{mgse}, we find that both the classical and gradient parts of the displacement field can be expressed through four independent functions each. Thus, the total number of independent functions that should be used to define CGP general solution is 8 that is similar to LBVB solution.

\subsection{Generalized Boussinesq-Galerkin (BG) solution}

Choosing appropriate definitions of Helmholtz potentials $\varphi $ and $%
\mathbf{A}$ \eqref{ht} in terms of a single vector function $\mathbf{G}$\textbf{, }an
extended form of the Boussinesq-Galerkin (BG) general solution of the
equilibrium equations \eqref{ee} was derived in the following form
\cite{solyaev2024complete}:
\begin{equation}
\mathbf{u}=\frac{1}{\alpha }\left( 1-l_{2}^{2}\nabla ^{2}\right) \nabla
 \nabla \cdot \mathbf{G} -\left( 1-l_{1}^{2}\nabla ^{2}\right)
\nabla \times \nabla \times \mathbf{G} 
\label{bggs1}
\end{equation}%
or equivalently%
\begin{equation}
\mathbf{u}=\left( 1-l_{1}^{2}\nabla ^{2}\right) \nabla ^{2}\mathbf{G}-\kappa
\left( 1-l_{3}^{2}\nabla ^{2}\right) \nabla\nabla \cdot \mathbf{G}
\label{bggs}
\end{equation}%
where $l_{3}^{2}=\frac{\alpha }{1-\alpha }\left( \frac{1}{\alpha }%
l_{2}^{2}-l_{1}^{2}\right)$
and $\mathbf{G}$ is the Galerkin vector stress function
that satisfies the following governing bi-Helmholtz/bi-Laplace equation:
\begin{equation}
\left( 1-l_{1}^{2}\nabla ^{2}\right) \left( 1-l_{2}^{2}\nabla ^{2}\right)
\nabla ^{2}\nabla ^{2}\mathbf{G}=-\frac{\mathbf{b}}{\mu }
\label{bggse}
\end{equation}

The BG representation is rather complex for solving particular boundary-value problems, as it requires the computation of high-order derivatives of $\textbf G$. However, it is convenient for proving the completeness of general solutions in SGE, by analogy with the classical approach \cite{gurtin1962helmholtz,solyaev2024complete}. It appears that a similar representation was first introduced in the work \cite{doyle1969general} within the third gradient elasticity theory. This work is rarely cited, and we only recently came across it, even though it also presents a generalization of Mindlin's solution \eqref{mgs} to the third-order theory.

\subsection{Generalized Papkovich-Neuber (PN) solution}

Based on the analysis of the structure of LBVB general solution \eqref{lgs}-\eqref{lgsge}, it was shown in Ref. \cite{solyaev2022elastic} that the number of potentials determining the gradient part of the displacement field can be reduced, and the representation \eqref{lgs} can be simplified. Subsequently, the resulting form of the general solution was derived in Ref. \cite{solyaev2024complete} based on deductive reasoning generalizing the classical Papkovich approach \cite{papkovich1932representation}. As a result, the simplest form of the Papkovich-Neuber (PN) general solution within SGE took the following form:
\begin{equation}
\begin{aligned}
\mathbf{u}&=\mathbf{u}_{c}+\mathbf{u}_{g},\\
\mathbf{u}_{c}&=\mathbf{B}%
_{c}-\tfrac{\kappa}{2} \nabla \left( \mathbf{r}\cdot \mathbf{B}_{c}+\phi
_{c}\right),\\
\mathbf{u}_{g}&=\mathbf{B}_{g}+l_{1}^{2}\nabla \phi
_{g},  
\label{pngs}
\end{aligned}
\end{equation}%
in which the stress functions $\mathbf{B}_{c},\mathbf{B}%
_{g},\phi _{c},\phi _{g}$ have to obey the following governing
equations:%
\begin{equation}
\begin{aligned}
\nabla ^{2}\mathbf{B}_{c} &=-\frac{\mathbf{b}}{\mu },   \\
\nabla ^{2}\phi _{c} &=\frac{\mathbf{r}\cdot \mathbf{b}}{\mu }%
+2l_{1}^{2}\left( 1-2\upsilon \right) \frac{\nabla \cdot \mathbf{b}}{\mu },
 \\
\left( 1-l_{2}^{2}\nabla ^{2}\right) \mathbf{B}_{g} &=-l_{2}^{2}\frac{\hat{\mathbf{b}}}{\mu },\qquad\nabla \cdot \mathbf{B}_{g}=0,   \\
\left( 1-l_{1}^{2}\nabla ^{2}\right) \phi _{g} &=
-l_{1}^{2}\frac{\nabla \cdot \mathbf{b}}{\alpha\mu }
\label{pnge}
\end{aligned}
\end{equation}
where $\hat{\mathbf{b}}$ is the rotational part of the body load (related, e.g., to Coriolis or ponderomotive forces)\footnote{The final solution representation in Ref. \cite{solyaev2024complete}  contains $\mathbf{b}$ instead of $\hat{\mathbf{b}}$. This is a typo, as the derivation clearly requires $\hat{\mathbf{b}}$ to maintain consistency with the right-hand side of the governing equation for $\textbf B_g$ (see Eqs. (41), (42) in \cite{solyaev2024complete}).}.

An advantage of this solution is that, similar to the LBVB representation, it explicitly separates the classical and gradient parts of the displacement field. The classical part $\textbf u_c$ coincides with the standard PN representation that is defined by $\mathbf{B}_c$ and $\phi
_{c}$. The gradient part is expressed through the minimal necessary number of additional potentials: a scalar potential $\phi _{g}$ for the irrotational part of $\textbf u_g$, and a solenoidal vector function $\mathbf{B}_{g}$ defining the rotational part of $\textbf u_g$. 
By using standard representations from the vector analysis \cite{morse1954methods}, solenoidal field $\mathbf{B}_{g}$ can be represented via two scalar potentials satisfying the Helmholtz equations (see, e.g. \cite{solyaev2022elastic, solyaev2024higher}). Therefore, in total we have only three independent functions that are used to define three components of $\textbf u_g$.
Moreover, the displacement representation in Eqs. \eqref{pngs}, \eqref{pnge} involves only first-order derivatives of the potentials. Owing to its simplicity, we refer to this specific form as the generalized PN representation within SGE throughout this work. 

Note that the representation for the gradient part of the displacement field, $\textbf{u}_g$ in \eqref{pngs}, \eqref{pnge}, can be viewed as a Helmholtz decomposition. This decomposition yields irrotational and solenoidal parts, $\nabla\phi_g$ and $\textbf{B}_g$, each satisfying a Helmholtz equation with a distinct length scale parameter ($l_1$ and $l_2$, respectively). Additional discussion of this decomposition is provided in Section 3.8.

Reduction of LBVB solution \eqref{lgs}-\eqref{lgsge} to the presented form of PN solution \eqref{pngs}, \eqref{pnge} was discussed in Ref. \cite{solyaev2022elastic}. Namely, it was shown that the relations between stress functions are the following:
\begin{equation}
\label{rellp}
\begin{aligned}
\textbf F_0 = \textbf B_c, \quad f = \phi_c, \quad
\textbf F_2 = \textbf B_g, \quad \nabla\cdot\textbf F_1 = -\phi_g
\end{aligned}
\end{equation}

The reduced form of PN representation \eqref{pngs}, \eqref{pnge} with $l_2=0$ and $\textbf B_g =0$ for the so-called dilatation gradient elasticity theory was considered in Ref. \cite{lurie2021dilatation}. The representation for the couple stress theory \cite{mindlin1962effects} can be obtained if we put $l_1=0$ and $\phi_g =0$ in \eqref{pngs}, \eqref{pnge}.

\subsection{Generalized Ter-Mkrtychan-Naghdi-Hsu (TNH) solution}

In classical elasticity, TNH solution employs an integral representation of the displacement field through a single vector function $\mathbf P$ \cite{ter1947general,naghdi1961representation}. Integration is performed over the body volume or over its surface (in the absence of body forces) \cite{naghdi1961representation, borodachev1987generalization}. An advantage of this representation is the fact that $\mathbf P$ is unique \cite{wang1985naghdi}. In classical elasticity, TNH representation can be generalized to a nonlinear formulation \cite{goodman1989use} and to media with spatially varying properties \cite{borodachev1987generalization}. 

At first, we present the derivation of the general TNH representation that extends an approach originally introduced within classical elasticity theory \cite{naghdi1961representation}. Following this approach in SGE it becomes possible to derive two distinct TNH-type representations. They differ in the definition of the vector potential, which is expressed via the  Green's function for the Helmholtz equation \eqref{ghe} with either parameter $l_1$ or $l_2$, responsible for gradient effects in the potential or solenoidal part of the displacement field, respectively. 

To obtain the first variant of generalized TNH solution within SGE, we represent the solution  of equilibrium equations \eqref{ee} according to the Heomholtz theorem \eqref{ht} and representation \eqref{P3}:
\begin{equation}
	\mathbf{u}=\mathbf{u}^{1}+\mathbf{u}^{2}, \qquad
	\mathbf{u}^{1}=\nabla \mathcal{N}\left( \nabla \cdot \mathbf{u}\right),
	\qquad\nabla \cdot \mathbf{u}^{2}=0
\label{htn}
\end{equation}
where we use notation $\mathbf{u}^{1}$ and $\textbf u^2$ for the potential and solenodial parts of $\textbf u$, respectively.


Substituting \eqref{htn} into \eqref{ee}, we obtain
\begin{equation}
\nabla ^{2}\left( \alpha \left( 1-l_{1}^{2}\nabla ^{2}\right) \mathbf{u}%
^{1}+\left( 1-l_{2}^{2}\nabla ^{2}\right) \mathbf{u}^{2}\right) =-\frac{%
\mathbf{b}}{\mu },  
\label{I4}
\end{equation}%
which can be written in the form%
\begin{equation}
\left( 1-l_{2}^{2}\nabla ^{2}\right) \nabla ^{2}\left( \alpha \mathcal{H}%
_{2}\left( 1-l_{1}^{2}\nabla ^{2}\right) \mathbf{u}^{1}+\mathbf{u}%
^{2}\right) =-\frac{\mathbf{b}}{\mu },  
\label{I5}
\end{equation}%
where $\mathcal{H}_{2}$ is the integral convolution operator \eqref{ghe} that corresponds to the Green's function of modified Helmholtz equation with the length scale parameter $l_{2}$, i.e. it is valid that $( 1-l_{2}^{2}\nabla ^{2})\mathcal{H}_{2}\textbf u^1 = \textbf u^1$. 

We next define the vector function within the brackets in \eqref{I5} by%
\begin{equation}
\mathbf{P}=\alpha \mathcal{H}_{2}\left( 1-l_{1}^{2}\nabla ^{2}\right)
\mathbf{u}^{1}+\mathbf{u}^{2} 
\label{I6}
\end{equation}%

This vector function $\textbf P$ should be treated as TNH potential within SGE. Combining (\ref{I5}) and (\ref{I6}), we get governing equation for this potential:%
\begin{equation}
\left( 1-l_{2}^{2}\nabla ^{2}\right) \nabla ^{2}\mathbf{P}=-\frac{\mathbf{b}%
}{\mu } 
\label{I7}
\end{equation}

Thus, according to \eqref{P1}, \eqref{bf}, within the first variant of TNH solution the representation for $\textbf P$ will contain the classical harmonic part that obeys the Poisson equation and the additional part that obeys the modified Helmholtz equation with the length scale parameter $l_2$.

Applying the Helmholtz theorem to $\mathbf{P}$, we can write
\begin{equation}
\mathbf{P}=\mathbf{P}^{1}+\mathbf{P}^{2}
  \label{I8}
\end{equation}
where $\mathbf{P}^{2}$ is its solenoidal part ($\nabla\cdot\mathbf{P}^{2}=0$) and for the potential part $\mathbf{P}^{1}$ we propose the following definition:
\begin{equation}
\mathbf{P}^{1}=\mathcal{H}_{2}\nabla \mathcal{N}\left( \left(
1-l_{2}^{2}\nabla ^{2}\right) \nabla \cdot \mathbf{P}\right)   
\label{I9}
\end{equation}

This definition extends the classical variant $\mathbf{P}^{1}=\nabla \mathcal{N}\left(  \nabla \cdot \mathbf{P}\right)$ \cite{naghdi1961representation}. We have selected this specific generalized form of $\textbf P^1$, which, upon substitution of \eqref{I6} into \eqref{I9}, yields the following useful result (see Appendix A):
\begin{equation}
\begin{aligned}
\mathbf{P}^{1} 
=\alpha \mathcal{H}_{2}\left( 1-l_{1}^{2}\nabla ^{2}\right) \mathbf{u}^{1}
\label{I10}
\end{aligned}
\end{equation}

Then, considering \eqref{I6}, \eqref{I8} and \eqref{I10} we find that:
\begin{equation}
\mathbf{P}^{2}=\mathbf{u}^{2}
\label{I12}
\end{equation}

It should be noted that the classical derivation of TNH solution leads to an analogous result \eqref{I12} (see \cite{naghdi1961representation}). From \eqref{I10}, we also find that
\begin{equation}
\begin{aligned}
\mathbf{u}^{1} &=\frac{1}{\alpha }\mathcal{H}_{1}\left( 1-l_{2}^{2}\nabla
^{2}\right) \mathbf{P}^{1}
\label{I11}
\end{aligned}
\end{equation}
where $\mathcal{H}_{1}$ is the integral convolution operator \eqref{ghe} that corresponds to the Green's function of the modified Helmholtz equation with the length scale parameter $l_{1}$.

Finally, using Eqs. (\ref{htn}), (\ref{I9}), (\ref{I12}) and (\ref{I11}), one can obtain the following form of generalized TNH solution within SGE (for derivations see Appendix A):
\begin{equation}
\mathbf{u} = \mathbf{P}-\kappa 
\nabla \left(
	\mathcal{N}\left( \nabla \cdot \mathbf{P}\right) -\tfrac{l_{2}^{2}-l_{1}^{2}}{1-\alpha }\,\mathcal{H}_{1}\left( \nabla \cdot \mathbf{%
P}\right) 
 \right)
\label{I13}
\end{equation}
where as stated above the TNH potential should obey the fourth-order governing equation \eqref{I7}.

An alternative form of TNH solution in SGE can be obtained if, in equation \eqref{I5}, we factor out the modified Helmholtz operator with the length scale parameter $l_1$. In this case, one can obtain the following representation (see Appendix A):
\begin{equation}
\mathbf{u} = \left( 1-\left( l_{1}^{2}-l_{2}^{2}\right) \nabla ^{2}\mathcal{H}%
_{2}\right) \mathbf{P}- \nabla \left(
\kappa\,\mathcal{N}\left( \nabla \cdot \mathbf{P}\right) -(l_{1}^{2}-l_{2}^{2}) \mathcal{H}_{2}\left( \nabla \cdot \mathbf{%
P}\right)
\right)
\label{I14}
\end{equation}
in which the governing equation for $\textbf P$ is the following:
\begin{equation}
\left( 1-l_{1}^{2}\nabla ^{2}\right) \nabla ^{2}\mathbf{P}=-\frac{\mathbf{b}%
}{\mu }.  
\label{I14g}
\end{equation}%

Note that classical TNH representation \cite{naghdi1961representation} follows from \eqref{I7}, \eqref{I13} as well as from \eqref{I14}, \eqref{I14g} under the standard assumption $l_1=l_2 = 0$:
\begin{eqnarray}
\mathbf{u} = \mathbf{P}-\kappa 
\nabla \left(
	\mathcal{N}\left( \nabla \cdot \mathbf{P}\right)
 \right), \qquad 
\nabla ^{2}\mathbf{P}=-\tfrac{\mathbf{b}}{\mu } 
\label{tnhc}
\end{eqnarray}

Taking into account \eqref{P1}, \eqref{bf}, it can be easily shown that TNH solution in SGE \eqref{I7}, \eqref{I13} or \eqref{I14}, \eqref{I14g} also admits an additive decomposition into a classical part of the displacement field defined by \eqref{tnhc} and an additional part related to gradient effects. It is also interesting to note that in the simplified SGE with $l_1=l_2=l$ the displacement representation formally reduces to classical relation, while remaining the higher-order governing equation for $\textbf P$:
\begin{equation}
\mathbf{u} = \mathbf{P}-\kappa 
\nabla \left(
	\mathcal{N}\left( \nabla \cdot \mathbf{P}\right)  \right),
	\qquad
	\left( 1-l^{2}\nabla ^{2}\right) \nabla ^{2}\mathbf{P}=-\tfrac{\mathbf{b}%
}{\mu }
\label{I13s}
\end{equation}

Presented forms of TNH displacement general solutions within SGE \eqref{I7}, \eqref{I13} and \eqref{I14}, \eqref{I14g} are new for the best of authors knowledge. We can also note that the integral representation for the displacement field in gradient elasticity theory was also addressed in \cite{dinh2024factorization} within the framework of quaternion analysis. Fundamental solutions and Green functions were examined within SGE, e.g. in Refs. \cite{lazar2018mindlin,gao2009green,gourgiotis2018concentrated}.

\subsection{Solution representation for specific boundary value problems}

In classical elasticity, there exist several general solutions for specialized formulations, such as those for problems without a rotation field (Lam\'e strain potential) and for axisymmetric problems (Love's solution and Boussinesq potentials). In this section, we demonstrate how these simplified representations can be derived from the general solutions of SGE presented above.

\subsubsection{Generalized Lam\'e strain potential}

Let us apply the Helmholtz decomposition \eqref{ht} to the displacement field in \eqref{ee} under the assumption of a zero vector potential $\textbf A =0$. This corresponds to a solution that contains no rotational part. The resulting representation for the displacement field and the governing equation for the scalar potential $\varphi$ -- referred to as the Lam\'e strain potential -- are then as follows:
\begin{equation}
\mathbf{u}=\nabla \varphi,\qquad  
\left( 1-l_{1}^{2}\nabla ^{2}\right) \nabla ^{2}\nabla \varphi =-\frac{\mathbf{b%
}}{\alpha \mu }
\label{lps}
\end{equation}
where the body load $\textbf b$ should also be a potential field.

The analysis and application of this representation for solving boundary value problems were discussed in detail in Ref. \cite{papargyri2015lame} within the simplified formulation with $l_1=l_2$. Note that in the general formulation of SGE, the influence of gradient effects may be determined not only by the value of the length scale parameter $l_1$ but also by other additional constants of SGE, due to influence of boundary conditions.

This representation \eqref{lps} can be also obtained as the particular case of the other  general solutions. For example, we can use $\textbf B=0$ in CGP solution \eqref{cgps} or $\textbf B_c =\textbf B_g =0$ in PN solution  \eqref{pngs}, etc.

\subsubsection{Generalized Love strain function}

Considering axisymmetric problems, we can assume that the Galerkin vector in \eqref{bggs} has a non-zero projection only along the symmetry axis, e.g. $\textbf G = G (r,z)\textbf e_z$, where $G(r,z)$ is the Love strain function defined in cylindrical coordinates $(r,\theta,z)$. In this case, we obtain the following form of the Love general solution within SGE:

\begin{equation}
\begin{aligned}
\mathbf{u}&=\left( 1-l_{1}^{2}\nabla ^{2}\right) \nabla ^{2}\mathbf{G}-\kappa
\left( 1-l_{3}^{2}\nabla ^{2}\right) \nabla \left( \nabla \cdot \mathbf{G}%
\right)  \\
&=\left( 1-l_{1}^{2}\nabla ^{2}\right) \nabla ^{2}\mathbf{G}-\kappa
\left( 1-l_{3}^{2}\nabla ^{2}\right) \nabla \left( \frac{\partial G}{\partial z}\right) 
\end{aligned}
\label{lsf}
\end{equation}
which in components becomes:
\begin{equation}
\begin{aligned}
u_r&=-\kappa
\left( 1-l_{3}^{2}\nabla ^{2}\right) 
\frac{\partial^2 G}{\partial r\partial z},\qquad u_\theta=0\\
u_z&=
\left( 1-l_{1}^{2}\nabla ^{2}\right) \nabla ^{2}G
-\kappa
\left( 1-l_{3}^{2}\nabla ^{2}\right) 
\frac{\partial^2 G}{\partial z^2} 
\end{aligned}
\label{222}
\end{equation}

The form of the governing equation for Galerkin vector \eqref{bggse} provides limitations for the cases, when the body load is acted along the axis of symmetry only. For the Love strain function it takes the form:

\begin{equation}
\left( 1-l_{1}^{2}\nabla ^{2}\right) \left( 1-l_{2}^{2}\nabla ^{2}\right)
\nabla ^{2}\nabla ^{2}G=-\frac{b_z}{\mu }
\label{lsfe}
\end{equation}

This representation reduces to classical form, when $l_1=l_2=0$. Its possible generalization for the problems with radial body forces can be obtained following approach suggested in Ref. \cite{simmonds2000love}, although the appropriate and more simple form of axisymmetric solution is provided by   Boussinesq potentials introduced in the next subsection.

\subsubsection{Generalized Boussinesq potentials}

For axisymmetric problems, Boussinesq proposed a simplified representation of the general solution, in which the order of derivatives of the potentials is reduced compared to Love's solution. Boussinesq axisymmetric solution can most easily be obtained as a special case of PN solution \eqref{pngs}, \eqref{pnge}, in which for classical part of displacement we should assume $\textbf B_c = \psi_c(r,z) \textbf e_z$, $\phi_{c}=\phi_{c}(r,z)$ and for gradient part we can use $\textbf B_g = \nabla\times (\psi_g(r,z)\textbf e_\theta)$, $\phi_{g}=\phi_{g}(r,z)$. Also we assume that the body load is the potential field $\textbf b = \nabla \phi_b(r,z)$. Then, from \eqref{pngs} we obtain:
\begin{equation}
\begin{aligned}
\mathbf{u}&=\mathbf{u}_{c}+\mathbf{u}_{g},\\
\mathbf{u}_{c}&=\psi_c \textbf e_z
-\tfrac{\kappa}{2} \nabla \varphi_c ,\\
\mathbf{u}_{g}&=\nabla\times (\psi_g \textbf e_\theta)+l_{1}^{2}\nabla \phi_{g},  
\label{bp}
\end{aligned}
\end{equation}
where we also introduce classical Boussinesq potential $\varphi_c = z \psi_c +\phi_{c}$ that is directly related to the scalar potential in standard Helmholtz decomposition \eqref{ht} applied to $\mathbf{u}_{c}$.

The components of the displacement field \eqref{bp} is given by:
\begin{equation}
\begin{aligned}
u_r&=
-\tfrac{\kappa}{2} \tfrac{\partial\varphi_c}{\partial r} 
-\tfrac{\partial\psi_g}{\partial z}
+l_{1}^{2}\tfrac{\partial\phi_g}{\partial r} \\
u_r&=
\psi_c
-\tfrac{\kappa}{2} \tfrac{\partial\varphi_c}{\partial z} 
+\tfrac{\partial\psi_g}{\partial r}
+\tfrac{\psi_g}{r}
+l_{1}^{2}\tfrac{\partial\phi_g}{\partial z} 
\label{bpc}
\end{aligned}
\end{equation}

From \eqref{pnge} it follows that the stress functions $\psi _c, \psi _g,\varphi_c,\phi _g$ have to obey the following governing
equations:%
\begin{equation}
\begin{aligned}
\nabla ^{2}\psi _c &=-\frac{1}{\mu }\tfrac{\partial \phi_b}{\partial z},   \\
  \nabla ^{2}\varphi _{c} &= 2\tfrac{\partial \psi_c}{\partial z}
  -\tfrac{z}{\mu } \tfrac{\partial \phi_b}{\partial z}
  + \tfrac{\mathbf{r}\cdot \nabla \phi_b}{\mu }%
+2l_{1}^{2}\left( 1-2\upsilon \right) \tfrac{\nabla^2 \phi_b}{\mu }
 \\
\left( 1-l_{2}^{2}\nabla ^{2}\right) \psi_{g} &=0\\
\left( 1-l_{1}^{2}\nabla ^{2}\right) \phi _{g} &=
-l_{1}^{2}\tfrac{\nabla^2 \phi_b}{\alpha\mu }
\label{bpe}
\end{aligned}
\end{equation}
where the equation for $\varphi_c$ is obtained by combining the equations for the classical PN potentials in \eqref{pnge}.

\subsection{General decomposition of displacement field into classical and gradient parts in SGE}

Let us consider the following modified form of generalized PN solution \eqref{pngs}: 

\begin{equation}
\begin{aligned}
\mathbf{u}&=\hat {\mathbf{u}}_{c}+\mathbf{u}^*_{c}+\mathbf{u}_{g},\\
\hat {\mathbf{u}}_{c}&=\mathbf{B}_{c}-\tfrac{\kappa}{2} \nabla \left( \mathbf{r}\cdot \mathbf{B}_{c}+\hat \phi_{c}\right)\\
\mathbf{u}^*_{c} &= -\tfrac{\kappa}{2} \nabla \phi^*_{c}\\
\mathbf{u}_{g}&=\mathbf{B}_{g}+l_{1}^{2}\nabla \phi
_{g} 
\label{pngsm}
\end{aligned}
\end{equation}%
where the stress functions $\mathbf{B}_{c},\mathbf{B}%
_{g},\phi _{g}$ have to obey the  governing equations that are given in \eqref{pnge}; while the classical scalar stress function is decomposed into two parts $\phi _c = \hat \phi_c + \phi_c^*$ that produce corresponding decomposition of classical part of displacement field $\textbf u = \hat {\mathbf{u}}_{c} + \mathbf{u}^*_{c}$. The introduced stress functions $\hat \phi_c,  \phi_c^*$ are used to separate the part of classical solution, which definition totally coincides with classical PN representation, including the right hand side in the governing equations for its potentials that take the form: 
\begin{equation}
\begin{aligned}
\nabla ^{2}\hat \phi _{c} &=\frac{\mathbf{r}\cdot \mathbf{b}}{\mu },
\qquad
 \nabla ^{2} \phi ^*_{c} &=
2l_{1}^{2}\left( 1-2\upsilon \right) \frac{\nabla \cdot \mathbf{b}}{\mu }
\label{phicx}
\end{aligned}
\end{equation}

Thus, from \eqref{pnge}, \eqref{pngsm}, \eqref{phicx}$_1$  we see that definition for $\hat {\textbf u}_c$ is the classical PN representation:
\begin{equation}
\begin{aligned}
&\hat {\mathbf{u}}_{c}=\mathbf{B}_{c}-\tfrac{\kappa}{2} \nabla \left( \mathbf{r}\cdot \mathbf{B}_{c}+\hat \phi_{c}\right)\\
&\nabla ^{2}\hat \phi _{c} =\frac{\mathbf{r}\cdot \mathbf{b}}{\mu },\quad
\nabla ^{2}\mathbf{B}_{c} =-\frac{\mathbf{b}}{\mu }
\label{hatu}
\end{aligned}
\end{equation}
 
Since the completeness of classical PN representation is proven \cite{wang2008recent}, then we can state that $\hat {\textbf u}_c$ is any kind of solution of equilibrium equations of classical elasticity, i.e. it is a solution of equation:
 \begin{equation}
\begin{aligned}
\alpha \nabla \nabla \cdot \hat{\mathbf{u}}_c
- \nabla \times \nabla \times \hat{\mathbf{u}}_c
=-\frac{\mathbf{b}}{\mu }
\label{eech}
\end{aligned}
\end{equation}
 
 Therefore, to define $\hat {\textbf u}_c$ we can use any kind of classical general solution that is known up to date (see, e.g. \cite{wang2008recent,liu2025newly}) and that is appropriate for the considered problem. At the same time, for the remaining introduced part of the solution $\textbf u^*_c$ we can just find appropriate particular solution that can be defined by using \eqref{phicx}$_2$, \eqref{P3} as follows:
 \begin{equation}
\begin{aligned}
\textbf u^*_{c} &=
-\tfrac{l_{1}^{2}}{\alpha\mu}
\nabla\varphi_b
\label{phix}
\end{aligned}
\end{equation}
where $\varphi_b$ is the scalar potential that defines the irrotational part of the body load so that  $\nabla^2\varphi_b = \nabla\cdot\textbf b$ (see \eqref{P3}, \eqref{eec}). 

Finally, we find that the general solution of equilibrium equations of SGE \eqref{ee} can be presented as follows:
\begin{equation}
\begin{aligned}
\mathbf{u}&=\mathbf{u}_{c}+\mathbf{u}_{g} 
-\tfrac{l_{1}^{2}}{\alpha\mu}\nabla\varphi_b
\label{gd}
\end{aligned}
\end{equation}
where $\mathbf{u}_{c}$ is any kind of general solution of classical linear isotropic elasticity including PN, TNH, BG or other (we avoid using hat symbol for this part of solution), $\varphi_b$ is the potential of body load, and $\textbf u_g$ is the gradient part of displacement field defined by Helmholtz decomposition according to \eqref{pngs}:
\begin{equation}
\mathbf{u}_{g}=\mathbf{B}_{g}+\nabla \phi_g, \qquad \nabla \cdot \mathbf{B}_{g}=0
\end{equation}
and its potentials should obey the following equations according to \eqref{pnge}:
\begin{equation}
\begin{aligned}
\left( 1-l_{2}^{2}\nabla ^{2}\right) \mathbf{B}_{g} 
		=-l_{2}^{2}\tfrac{\hat{\mathbf{b}}}{\mu}, \qquad
\left( 1-l_{1}^{2}\nabla ^{2}\right) \phi _{g} =
-l_{1}^{4}\tfrac{\nabla \cdot \mathbf{b}}{\alpha\mu }
\label{gdeg}
\end{aligned}
\end{equation}

Representation \eqref{gd}-\eqref{gdeg}  can be considered as the extension of the results presented by Charalambopoulos et al. \cite{charalambopoulos2022representing} for simplified SGE (with $l_1=l_2$) to the general Toupin-Mindlin SGE (with $l_1\neq l_2$) in the presence of arbitrary body load. Related discussion on the decomposition of SGE general solution into classical and gradient parts was also given in Refs. \cite{lurie2006interphase,lurie2009advanced, lurie2011eshelby, lurie2023general,lurie2024general,volkov2010eshelby}.
We can also note the so-called Ru-Aifantis theorem in simplified SGE \cite{ru1993simple,askes2011gradient}, which provides a decomposition of the displacement field into interrelated gradient and classical parts. However, this theorem can be applied only in unbounded homogeneous domains \cite{gutkin1999dislocations, lazar2006dislocations, lazar2015non}.

In some simple cases the representation \eqref{gd}-\eqref{gdeg} can be directly used for derivation of analytical solutions within SGE based on the known closed-form solutions of classical elasticity $\mathbf{u}_c$, although there exist many examples, when the coefficients of classical part of general solution are modified within SGE in comparison to classical solutions \cite{eshel1970effects, khakalo2017gradient, solyaev2019three}. Moreover, there exist examples of boundary value problems, for which the closed-form classical solution cannot be reproduced in SGE and infinite series containing coupled classical and gradient terms arise (see, e.g. \cite{lomakin2020stress}). 
Nevertheless, the representation given by Eqs. \eqref{gd}-\eqref{gdeg} can be useful in applications, as it enables the use of any classical general solution representation and provides a simple definition for the gradient part of the displacement field. It is also useful from a theoretical standpoint, for example, for testing the structure and correctness of analytical solutions within SGE.

\section{Relations between stress functions}

In order to derive the completeness theorem for any of the general solutions to SGE equilibrium equations \eqref{ee} that were discussed above, we need to find the relations between  different stress functions of these solutions. Also, it is important to show the equivalence of different forms of general solutions developed independently within SGE. 

Note that the equivalence of the potentials in Mindlin solution (Section 3.1) and in CGP solution (section 3.3) was shown in Ref. \cite{charalambopoulos2020plane}. The variants of general solutions for special problems discussed in Section 3.7 are shown to be the particular cases of BG, PN and other solutions, so we also do not need to consider these special solutions separaterly. Relation between PN solution and LBVB solution was discussed in details in Ref. \cite{solyaev2022elastic} and relations between their potentials are given by \eqref{rellp}.

Therefore, we need to consider only the relations between BG and PN solutions, BG and TNH solutions, and PN and TNH solutions within SGE. Finally, we will show the relations between PN and Mindlin solutions that will complete the task since it will automatically provide the interrelations for LVBV and CGP solutions.

\subsection{BG and PN potentials}

Let us begin with the relations between BG and PN stress functions.  PN stress functions $\mathbf{B}_{c},\mathbf{B}_{g},\varphi _{c},\varphi _{g}$ \eqref{pngs} in terms of the Galerkin stress function $\mathbf{G}$ \eqref{bggs1} were obtained in the
following form \cite{solyaev2024complete}:
\begin{equation}
\begin{aligned}
\mathbf{B}_{c} &=\left( 1-l_{1}^{2}\nabla ^{2}\right) \left(
1-l_{2}^{2}\nabla ^{2}\right) \nabla ^{2}\mathbf{G} \\
\mathbf{B}_{g} &=-l_{2}^{2}\left( 1-l_{1}^{2}\nabla ^{2}\right) \nabla
^{2}\left( \nabla \times \nabla \times \mathbf{G}\right) \\
\varphi _{c} &=2\left( 1-l_{1}^{2}\nabla ^{2}\right) \left(
1-l_{3}^{2}\nabla ^{2}\right) \nabla \cdot \mathbf{G}-\mathbf{r}\cdot
\mathbf{B}_{c}+2l_{1}^{2}\nabla \cdot \mathbf{B}_{c} \\
\varphi _{g} &=l_{1}^{2}\nabla ^{2}\left( 1-l_{3}^{2}\nabla ^{2}\right)
\nabla \cdot \mathbf{G}-l_{1}^{2}\nabla \cdot \mathbf{B}_{c} 
\label{bgpn1}
\end{aligned}
\end{equation}

The inverse relation was not discussed previously, although we can find it using \eqref{bggs}, \eqref{pngs} and \eqref{bgpn1}. Namely, assuming the equality of the displacements as defined in the BG \eqref{bggs} and PN \eqref{pngs} solutions, we can obtain:
\begin{equation}
\begin{aligned}
	\left( 1-l_{1}^{2}\nabla ^{2}\right) \nabla ^{2}\mathbf{G}&=
\mathbf{B}_{c}-\tfrac{\kappa}{2} \nabla \left( \mathbf{r}\cdot \mathbf{B}_{c}+\phi_{c}\right)\\
&+\mathbf{B}_{g}+l_{1}^{2}\nabla \phi_{g}
+\kappa
\left( 1-l_{3}^{2}\nabla ^{2}\right) \nabla \nabla \cdot \mathbf{G} 
\label{bgpn2}
\end{aligned}
\end{equation}

To define the last term in the RHS of \eqref{bgpn2} we can use the particular solution for $\textbf G$ in \eqref{bgpn1}$_1$. Therefore, we have:
\begin{equation}
\begin{aligned}
	\left( 1-l_{1}^{2}\nabla ^{2}\right) \nabla ^{2}\mathbf{G}
	&=
\mathbf{B}_{c}+\mathbf{B}_{g}-\tfrac{\kappa}{2} \nabla \Big( \mathbf{r}\cdot \mathbf{B}_{c}+\phi_{c}
	-\tfrac{2l_{1}^{2}}{\kappa}\nabla \phi_{g} \\
&-2
( 1-l_{3}^{2}\nabla ^{2}) \nabla \cdot \mathcal{N}(\mathcal{H}_{2}(\mathcal{H}_{1}(\mathbf{B}_{c})))
\Big) 
\label{bgpn3}
\end{aligned}
\end{equation}

Finally, the Galerkin stress function $\mathbf{G}$ can be presented in terms of PN potentials as follows:
\begin{equation}
\begin{aligned}
	\mathbf{G}
	&=
\mathcal{N}\Big[\mathcal{H}_{1}\Big[
\mathbf{B}_{c}+\mathbf{B}_{g}-\tfrac{\kappa}{2} \nabla \Big( \mathbf{r}\cdot \mathbf{B}_{c}+\phi_{c}
	-\tfrac{2l_{1}^{2}}{\kappa}\nabla \phi_{g} \\
&-2
( 1-l_{3}^{2}\nabla ^{2}) \nabla \cdot \mathcal{N}(\mathcal{H}_{2}(\mathcal{H}_{1}(\mathbf{B}_{c})))
\Big]\Big]
\label{bgpn4}
\end{aligned}
\end{equation}

Note that given relations \eqref{bgpn1}, \eqref{bgpn4} are reduced to those of classical elasticity \cite{wang2008recent} when $l_1=l_2=l_3=0$ (meaning also that $\textbf B_g=0$, $\phi_g=0$, $\mathcal{H}_{1}(\mathcal B) = \mathcal{H}_{2}(\mathcal B) = \mathcal B$):
\begin{equation}
\begin{aligned}
\mathbf{B}_{c} &=\nabla ^{2}\mathbf{G}, \qquad
\varphi _{c} =
2\nabla \cdot \mathbf{G}-\mathbf{r}\cdot(\nabla ^{2}\mathbf{G})\\[5pt]
	\mathbf{G}
	&=
\mathcal{N}(\mathbf{B}_{c})-\tfrac{\kappa}{2} \mathcal{N}\nabla \left( \mathbf{r}\cdot \mathbf{B}_{c}+\phi_{c}
-2\nabla \cdot \mathcal{N}(\mathbf{B}_{c})
\right)
\label{bgpn5}
\end{aligned}
\end{equation}

\subsection{BG and TNH potentials}

The relation between BG and TNH potentials can be obtained by extension of the approach that was suggested within the classical elasticity in Ref. \cite{wang1985naghdi}, although this extension is not trivial.

Let us introduce the following auxiliary vector field that is related to TNH potential $\textbf P$ as follows:
\begin{equation}
\begin{aligned}
&\mathbf{M} = \mathcal G(\textbf P), \qquad 
(1-l_1^2\nabla^2)\nabla^2\mathbf{M} = \textbf P
\label{bgtnh1}
\end{aligned}
\end{equation}
where $\mathcal G(\textbf P)$ defines the integral convolution of TNH potential $\textbf P$ with the Green's function of operator $(1-l_1^2\nabla^2)\nabla^2$. Based on the representation of the solutions of this operator \eqref{P1}, \eqref{bf} we can define $\mathcal G(\textbf P)$ in terms of corresponding integral operators of Laplace equation \eqref{gle} and modified Helmholtz equation \eqref{ghe} as follows:
\begin{equation}
\begin{aligned}
\mathcal G(\textbf P) =  \mathcal N(\textbf P) + l_1^2\mathcal H_1(\textbf P)
\label{bgtnh2}
\end{aligned}
\end{equation}
and the following standard relations for the particular solutions defined via the Green's functions are valid:
\begin{equation}
\begin{aligned}	
\nabla^2 \mathcal N(\textbf P) = \textbf P,
\qquad (1-l_1^2\nabla^2) \mathcal H_1(\textbf P) = \textbf P
\label{nh1p}
\end{aligned}
\end{equation}

Next, we observe, that the following identity is fulfilled:
\begin{equation}
\begin{aligned}
\mathcal G(\nabla\cdot\textbf P) = \nabla\cdot\mathcal G(\textbf P) + M_0
\implies 
\nabla\cdot\textbf M = \mathcal G(\nabla\cdot\textbf P) - M_0
\label{bgtnh3}
\end{aligned}
\end{equation}
in which the scalar function $M_0$ defines the commutator term for the integral operator $\mathcal G$ and the divergence operator $(\nabla\cdot)$.

Note that in classical elasticity $M_0$ satisfies the Laplace equation, however, in SGE to provide \eqref{bgtnh3} it should obey the following fourth-order equation:
\begin{equation}
\begin{aligned}
(1-l_1^2\nabla^2)\nabla^2M_0 = 0
\label{bgtnh4}
\end{aligned}
\end{equation}

Let us introduce then the definition for Galerkin stress function $\textbf G$:
\begin{equation}
\begin{aligned}
\textbf G = \textbf M 
- c_1 (1-l_1^2\nabla^2)\nabla\mathcal G(M_0)
+ c_2 \nabla M_0
\label{bgtnh5}
\end{aligned}
\end{equation}
where $c_1$ and $c_2$ are the constants to be determined from the following analysis.

The suggested definition \eqref{bgtnh5} generalizes the classical one 
$\textbf G = \textbf M 
- c_1\nabla\mathcal N(M_0)$ ($c_1 = 1/(1-2\nu)$) \cite{wang1985naghdi}. 
In this expression \eqref{bgtnh5}, the form of the first terms is chosen in a rather straightforward manner, by analogy with the classical representation, whereas the third additional term ($\nabla M_0$) has to be introduced specifically to eliminate the unknown function $M_0$ from the displacement representation, which will be obtained below. The specific form of this term represents our proposed approach for proving the relationship between the BG and TNH potentials. Although, other possible ways for this proof may exist. Namely, within the simplified SGE ($l_1=l_2$) one can use $c_2=0$ in \eqref{bgtnh5}.

Based on equations \eqref{bgtnh4}, \eqref{bgtnh5} we can also find:
\begin{equation}
\begin{aligned}
\nabla\cdot\textbf G &= \nabla\cdot\textbf M - c_1 M_0+ c_2 \nabla^2 M_0\\
(1-l_1^2\nabla^2)\nabla^2\textbf G &= \textbf P 
- c_1 (1-l_1^2\nabla^2)\nabla M_0
\label{bgtnh6}
\end{aligned}
\end{equation}

From \eqref{bgtnh6}, it follows that the introduced relationship between $\textbf G$ and $\textbf P$ ensures a proper connection between the governing equations imposed on these potentials. In particular, if we assume that \eqref{bggse} holds for $\textbf G$, then \eqref{bgtnh4}, \eqref{bgtnh6} imply \eqref{I7} for $\textbf P$. The converse is also true, so we can assert that relation \eqref{bgtnh5} together with \eqref{bgtnh4} guarantees the equivalence of governing equations \eqref{bggse} and \eqref{bgtnh6}:
\begin{equation}
\begin{aligned}
(1-l_1^2\nabla^2)(1-l_2^2\nabla^2)\nabla^2\nabla^2\textbf G =- \tfrac{\textbf b}{\mu} \quad\Leftrightarrow \quad
(1-l_2^2\nabla^2)\nabla^2\textbf P =- \tfrac{\textbf b}{\mu}
\label{bgtnh7}
\end{aligned}
\end{equation}

Moreover, we can show that the displacement representations of BG and TNH (first variant) solutions can be also related to each other by using \eqref{bgtnh5}.  Namely, from this relation we can find $c_1,c_2$ in \eqref{bgtnh5}. Thus, let us assume that BG solution \eqref{bggs} hold. Then, using \eqref{bgtnh3}, \eqref{bgtnh6} in \eqref{bggs} we can define the displacement field as follows:
\begin{equation}
\begin{aligned}
\textbf u &= 
(1-l_1^2\nabla^2)\nabla^2\textbf G 
- \kappa (1-l_3^2\nabla^2)\nabla\nabla\cdot\textbf G \\
&=\textbf P-c_1(1-l_1^2\nabla^2)\nabla M_0
- \kappa (1-l_3^2\nabla^2)\nabla(\nabla\cdot\textbf M - c_1 M_0+ c_2 \nabla^2 M_0)
\\&=\textbf P-\kappa(1-l_3^2\nabla^2)\nabla(\mathcal G(\nabla\cdot\textbf P))
-(c_1 - \kappa - c_1\kappa)\nabla M_0\\&
+(c_1 l_1^2-\kappa l_3^2-c_1\kappa  l_3^2-c_2\kappa (l_1^2-l_3^2)l_1^{-2})\nabla^2 \nabla M_0
\label{bgtnh8}
\end{aligned}
\end{equation}
where we also take into account that $\nabla^2 \nabla^2 M_0 = l_1^{-2}\nabla^2 M_0$ according to \eqref{bgtnh4}. 

From \eqref{bgtnh8}, it is seen that the terms related to the unknown arbitrary function $M_0$ can be avoided in the definition of $\textbf u$ by the appropriate choice of the constants $c_1$ and $c_2$. Namely, the following definition provide zero values of the last two bracketed terms in \eqref{bgtnh8}:
\begin{equation}
\begin{aligned}
c_1 = \frac{\kappa}{1-\kappa} = \frac{1}{1-2\nu}, \qquad
c_2 = \frac{l_1^2}{1-\kappa} = \frac{2(1-\nu)}{1-2\nu}l_1^2
\label{bgtnh9}
\end{aligned}
\end{equation}

Then, using \eqref{bgtnh2}, \eqref{bgtnh9} in \eqref{bgtnh8}, we find:
\begin{equation}
\begin{aligned}
\textbf u 
&= \textbf P
-\kappa(1-l_3^2\nabla^2)\nabla
\big(
\mathcal N(\nabla\cdot\textbf P)
+ l_1^2\mathcal H_1(\nabla\cdot\textbf P)
\big)\\
&= \textbf P
-\kappa\nabla
\big(
\mathcal N(\nabla\cdot\textbf P) - l_3^2\nabla\cdot\textbf P
+ l_1^2\nabla\cdot\textbf P
+ (l_1^2-l_3^2)l_1^2\nabla^2\mathcal H_1(\nabla\cdot\textbf P)\\
%
%
&= \textbf P
-\kappa\nabla
\big(
\mathcal N(\nabla\cdot\textbf P) 
+ (l_1^2-l_3^2)\mathcal H_1(\nabla\cdot\textbf P)
\big)
\label{bgtnh10}
\end{aligned}
\end{equation}
where we take into account that $\nabla^2\mathcal H_1(\nabla\cdot\textbf P) = l_1^{-2}(\mathcal H_1(\nabla\cdot\textbf P) - \nabla\cdot\textbf P)$ according to \eqref{nh1p}.

The resulting representation in \eqref{bgtnh10} coincides with the first variant of PNH solution \eqref{I13} if we take into account definition for $l_3$ in terms of $l_1$ and $l_2$ (see \eqref{bggs}). In a similar way one can derive  BG solution, assuming that TNH solution hold. Also, following similar approach the relation between BG solution and the second variant of TNH solution \eqref{I14}, \eqref{I14g} can be proved.

The found relation between BG and TNH potentials within SGE is given by \eqref{bgtnh5}. This relation contains some arbitrary function $M_0$ that obey the fourth-order equation \eqref{bgtnh4}. This function can be defined based on relation \eqref{bgtnh3} in terms of TNH potential $\textbf P$. In a similar way, we can define this function $M_0$ in terms of BG potential $\textbf G$ taking into account the following identity:
\begin{equation}
\begin{aligned}
\nabla\cdot \textbf G = \mathcal G\Big(\nabla\cdot
\big((1-l_1^2\nabla^2)\nabla^2 \textbf G\big)\Big)+ M_0
\label{bgtnh11}
\end{aligned}
\end{equation}
where $M_0$ is also some arbitrary function that should obey equation \eqref{bgtnh4}.

Thus, based on definitions \eqref{bgtnh5}, in which we should take into account \eqref{bgtnh1}, \eqref{bgtnh9} and identities \eqref{bgtnh3}, \eqref{bgtnh11}, we have the following relations between TNH and BG potentials:
\begin{equation}
\begin{aligned}
\textbf G &= \mathcal G(\textbf P) 
- \tfrac{1}{1-2\nu} (1-l_1^2\nabla^2)\nabla\mathcal G(\mathcal G(\nabla\cdot\textbf P) - \nabla\cdot\mathcal G(\textbf P))\\&
+\tfrac{2(1-\nu)}{1-2\nu}l_1^2 \nabla (\mathcal G(\nabla\cdot\textbf P) - \nabla\cdot\mathcal G(\textbf P))\\[5pt]
\textbf P &= (1-l_1^2\nabla^2)\nabla^2\textbf G\\&
+ \tfrac{1}{1-2\nu} (1-l_1^2\nabla^2)\nabla\mathcal G(\nabla\cdot \textbf G - \mathcal G(\nabla\cdot
\big((1-l_1^2\nabla^2)\nabla^2 \textbf G\big)))\\&
-\tfrac{2(1-\nu)}{1-2\nu}l_1^2 \nabla (\nabla\cdot \textbf G - \mathcal G(\nabla\cdot
\big((1-l_1^2\nabla^2)\nabla^2 \textbf G\big)))
\label{bgtnh12}
\end{aligned}
\end{equation}

The equations for the potentials are quite complicated. However, they are essential to demonstrate the completeness of TNH solution. The corresponding classical relations  \cite{wang2008recent} can be recovered from \eqref{bgtnh12} under assumption $l_1=l_2=0$:
\begin{equation}
\begin{aligned}
\textbf G &= \mathcal N(\textbf P) 
- \tfrac{1}{1-2\nu} \nabla\mathcal N(\mathcal N(\nabla\cdot\textbf P) - \nabla\cdot\mathcal N(\textbf P))\\[5pt]
\textbf P &= \nabla^2\textbf G
+ \tfrac{1}{1-2\nu} \nabla\mathcal N(\nabla\cdot \textbf G - \mathcal N(\nabla\cdot
\big(\nabla^2 \textbf G\big)))
\label{bgtnh13}
\end{aligned}
\end{equation}

\subsection{PN and TNH potentials}

The relationship between the PN and TNH potentials in classical elasticity theory was proposed in works \cite{tran1981notes, wang1985naghdi}. By generalizing this approach, a similar relationship in SGE can also be found. However, it seems simpler to adopt an approach where we utilize the already established connections between the PN and BG potentials (Section 4.1) and between the TNH and BG potentials (Section 4.2). Therefore, substituting \eqref{bgtnh12}$_1$ into \eqref{bgpn1} one can obtain definitions for PN vector and scalar potentials in terms of TNH potential. The inverse relation can be obtained by substitution \eqref{bgpn4} into \eqref{bgtnh12}$_2$. The resulting expressions become rather lengthy and are therefore omitted here.

\subsection{Relation between Mindlin's solution and generalized PN solution}

Let us now derive the relations between the Mindlin and the PN stress functions by reducing the Mindlin representation (\ref{mgs}) to
the generalized PN representation (\ref{pngs}). It follows from representation \eqref{P1}, \eqref{bf} and the governing equation for the vector potential $\mathbf{B}$ (\ref{mgse})$_{1}$ that%
\begin{equation}
\mathbf{B}=\mathbf{B}_{c}+\mathbf{B}_{g},  
\label{G11}
\end{equation}
where the stress functions $\mathbf{B}_{c}$ and $\mathbf{B}_{g}$ have to satisfy the equations
\begin{equation}
\nabla ^{2}\mathbf{B}_{c}=-\frac{\mathbf{b}}{\mu },\qquad
\left(
1-l_{2}^{2}\nabla ^{2}\right) \mathbf{B}_{g}=-l_{2}^{2}\frac{\mathbf{b}}{\mu
},  \label{G12}
\end{equation}%
and from \eqref{bfc}, we also have
\begin{equation}
\mathbf{B}_{c}=\left( 1-l_{2}^{2}\nabla ^{2}\right) \mathbf{B},\qquad
\mathbf{B}_{g}=l_{2}^{2}\nabla ^{2}\mathbf{B}.  
\label{G13}
\end{equation}

Thus, we immediately establish the direct and inverse relationships between the vector potentials of Mindlin’s solution and those of the generalized PN solution. It remains only to show that the gradient potential $\textbf B_g$ can be a purely solenoidal field, which will be demonstrated later.
We next use representation \eqref{P1}, \eqref{bf} for the governing equation for the potential $B_0$ \eqref{mgse}$_{2}$ to obtain
\begin{equation}
	B_0=B_{c}+B_{g}
	\label{G14}
\end{equation}
in which the stress functions $B_{c}$ and $B_{g}$ have to obey the following equations:%
\begin{equation}
\begin{aligned}
\nabla ^{2}B_{c} &=\frac{1}{\mu }\left( \mathbf{r}\cdot
\left( 1-l_{1}^{2}\nabla ^{2}\right) \mathbf{b}-4l_{1}^{2}\nabla \cdot
\mathbf{b}\right) ,   \\
\left( 1-l_{1}^{2}\nabla ^{2}\right) B_g &=\frac{
l_{1}^{2}}{\mu }\left( \mathbf{r}\cdot \left( 1-l_{1}^{2}\nabla ^{2}\right)
\mathbf{b}-4l_{1}^{2}\nabla \cdot \mathbf{b}\right)
\label{G15}
\end{aligned}
\end{equation}
and according to Eq. \eqref{bfc}, we also have
\begin{equation}
B_{c}=\left( 1-l_{1}^{2}\nabla ^{2}\right) B_0,\qquad
B_{g}=l_{1}^{2}\nabla ^{2}B_0
  \label{G16}
\end{equation}

The potentials $B_c$, $B_g$ cannot be considered as the scalar potentials of the PN solution $\phi_c$, $\phi_g$, since they satisfy equations with different right-hand sides (cf. \eqref{G15} and \eqref{pnge}). In order to introduce the PN potentials, we will perform the following additional transformations.
Employing Eq. (\ref{G15})$_{1}$ and the identity $\nabla ^{2}\left( \mathbf{r%
}\cdot \mathbf{b}\right) =\mathbf{r}\cdot \nabla ^{2}\mathbf{b}+2\nabla
\cdot \mathbf{b}$, we get
\begin{equation*}
\nabla ^{2}B_{c}=\frac{\mathbf{r}\cdot \mathbf{b}}{\mu }-%
\frac{l_{1}^{2}}{\mu }\nabla ^{2}\left( \mathbf{r}\cdot \mathbf{b}\right)
-\frac{2l_{1}^{2}}{\mu }\nabla \cdot \mathbf{b},
\end{equation*}
and thus we find
\begin{equation}
\nabla ^{2}\left(B_{c}+\frac{l_{1}^{2}}{\mu}\,\mathbf{r}\cdot
\mathbf{b}\right) =\frac{\mathbf{r}\cdot \mathbf{b}}{\mu }-2l_{1}^{2}%
\frac{\nabla \cdot \mathbf{b}}{\mu }.  \label{G17}
\end{equation}

Based on this relation \eqref{G17} and using \eqref{G13}, \eqref{G16}, we can define the scalar potential of generalized PN solution $\phi_{c}$ in terms of Mindlin's potentials as follows:%
\begin{equation}
\begin{aligned}
	\phi_{c}&=
	B_{c}
	+\frac{l_{1}^{2}}{\mu }\mathbf{r}\cdot \mathbf{b}
	-4l_{1}^{2}( 1-\upsilon) \nabla \cdot \mathbf{B}_{c}\\
	&=
	\left( 1-l_{1}^{2}\nabla ^{2}\right) B_0
	+\frac{l_{1}^{2}}{\mu }\mathbf{r}\cdot \mathbf{b}
	-4l_{1}^{2}( 1-\upsilon) \left( 1-l_{2}^{2}\nabla ^{2}\right) (\nabla \cdot \mathbf{B})
\end{aligned}
\label{G18}
\end{equation}%
so that using this relation in (\ref{G17}) and taking into account (\ref{G12})$_{1}$  we can obtain the correct
governing equation for $\phi _{c}$ (see \eqref{pnge}$_2$):
\begin{equation}
\nabla ^{2}\phi _{c}=\frac{\mathbf{r}\cdot \mathbf{b}}{\mu }%
+2l_{1}^{2}\left( 1-2\upsilon \right) \frac{\nabla \cdot \mathbf{b}}{\mu }
\label{G19}
\end{equation}

In a similar way, we can obtain from Eq. (\ref{G15})$_{2}$:
\begin{equation*}
\left( 1-l_{1}^{2}\nabla ^{2}\right) B_{g}=\frac{l_{1}^{2}%
}{\mu }\left( 1-l_{1}^{2}\nabla ^{2}\right) \mathbf{r}\cdot \mathbf{b}%
-\frac{2l_{1}^{4}}{\mu}\nabla \cdot \mathbf{b}
\end{equation*}%
that is equivalent to
\begin{equation}
\left( 1-l_{1}^{2}\nabla ^{2}\right) \left( B_{g}
-\frac{l_{1}^{2}}{\mu }\mathbf{r}\cdot \mathbf{b}\right) =-2l_{1}^{4}%
\frac{\nabla \cdot \mathbf{b}}{\mu }  
\label{G20}
\end{equation}

Therefore, let us introduce the following definition for scalar gradient potential of generalized PN solution:%
\begin{equation}
\begin{aligned}
	\phi _{g}&=\frac{1}{2\alpha l_{1}^{2}}\left(
	B_{g}-\frac{l_{1}^{2}}{\mu }\mathbf{r}\cdot \mathbf{b}%
	\right)	
	=\frac{1}{2\alpha }\left(
	\nabla ^{2}B_0-\frac{1}{\mu }\mathbf{r}\cdot \mathbf{b}%
	\right)
\end{aligned}   
\label{G21}
\end{equation}%
then using this relation in Eq. (\ref{G20}), the governing equation for PN potential $\phi _{g}$ takes the correct form (see \eqref{pnge}$_4$):
\begin{equation}
\left( 1-l_{1}^{2}\nabla ^{2}\right) \varphi _{g}=
-l_{1}^{2}\frac{\nabla \cdot \mathbf{b}}{\alpha\mu }
 \label{G22}
\end{equation}

Thus, we find the representations of PN scalar potentials in terms of Mindling potentials \eqref{G18}, \eqref{G21}. Inverse relation for scalar Mindlin potential can be also easily find by using \eqref{G14}, \eqref{G18}, \eqref{G21} as follows:
\begin{equation}
	B_0=\phi_{c}+2\alpha l_{1}^{2}\phi_{g}
	+4l_{1}^{2}( 1-\upsilon) \nabla \cdot \mathbf{B}_{c}
	\label{Gb0}
\end{equation}

The remaining task is to show that the representation of displacement field in Mindlin's solution \eqref{mgs} and in the generalized PN solution \eqref{pngs} are the same, if we take into account established interrelations between their potentials.
Substituting (\ref{G11}), (\ref{G12}) into \eqref{mgs}, we find that%
\begin{equation*}
\mathbf{u}=\mathbf{B}_{c}+\mathbf{B}_{g}-l_{2}^{2}\nabla \nabla \cdot
\mathbf{B}_{c}-l_{2}^{2}\nabla \nabla \cdot \mathbf{B}_{g}
-\tfrac{1}{2}\left(
\kappa -l_{1}^{2}\nabla ^{2}\right) \nabla \left( \mathbf{r}\cdot \mathbf{B}%
_{c}+B_0\right) 
\end{equation*}%

By using identity $\nabla ^{2}\left( \mathbf{r}\cdot \mathbf{B}%
_{c}\right) =\mathbf{r}\cdot \nabla ^{2}\mathbf{B}_{c}+2\nabla \cdot \mathbf{%
B}_{c}$, the previous equation becomes to:%
\begin{equation}
\begin{aligned}
\mathbf{u} &=
\mathbf{B}_{c}+\mathbf{B}_{g}
-l_{2}^{2}\nabla \nabla \cdot\mathbf{B}_{c}
-l_{2}^{2}\nabla \nabla \cdot \mathbf{B}_{g}\\
&-\tfrac{1}{2}\nabla
\left(
\kappa \,\mathbf{r}\cdot \mathbf{B}_{c} 
+ \tfrac{l_{1}^{2}}{\mu }\mathbf{r}\cdot \mathbf{b}
- 2l_{1}^{2}\nabla \cdot \mathbf{B}_{c}+\left( \kappa
- l_{1}^{2}\nabla ^{2}\right) B_0
\right)
\label{G23}
\end{aligned}
\end{equation}%
where we take into account (\ref{G12})$_{1}$.

Let us introduce the Helmholtz decomposition for the vector potential $\mathbf{B}_{g}$: 
\begin{equation}
\mathbf{B}_{g}=\nabla \Psi +\nabla \times \mathbf{\Psi},\qquad\nabla
\cdot \mathbf{B}_{g}=\nabla ^{2}\Psi ,\qquad
\nabla \times \mathbf{B}_{g}=-\nabla ^{2}\mathbf{\Psi}
\label{G24}
\end{equation}%
where the introduced scalar potential $\Psi $ and the solenoidal vector potential $%
\mathbf{\Psi}$ have to satisfy the following equations (that follows from (\ref{G12})$_{2}$):%
\begin{equation}
\left( 1-l_{2}^{2}\nabla ^{2}\right) \Psi =-l_{2}^{2}\frac{%
\nabla \cdot \mathbf{b}}{\mu },\qquad\left( 1-l_{2}^{2}\nabla ^{2}\right)
\mathbf{\Psi }=l_{2}^{2}\frac{\nabla \times \mathbf{b}}{\mu }
\label{G25}
\end{equation}%

Substituting (\ref{G24}) into (\ref{G23}), we obtain
\begin{equation}
\begin{aligned}
\mathbf{u} &=
\mathbf{B}_{c}+\nabla\times\mathbf\Psi
-\nabla^2 
\left(
l_{2}^{2}\nabla \cdot \mathbf{B}_{c}
-(1-l_{2}^{2}\nabla ^{2})\Psi 
\right) \\
&-\tfrac{1}{2}\nabla
\left(
\kappa \,\mathbf{r}\cdot \mathbf{B}_{c} 
+ \tfrac{l_{1}^{2}}{\mu }\mathbf{r}\cdot \mathbf{b}
- 2l_{1}^{2}\nabla \cdot \mathbf{B}_{c}+\left( \kappa
- l_{1}^{2}\nabla ^{2}\right) B_0
\right)
\label{G26}
\end{aligned}
\end{equation}
%

From the governing equations (\ref{G12})$_{1}$ and (\ref{G25})$_{1}$ it follows that the the term with Laplace operator in \eqref{G26} identically equals to zero:
\begin{equation*}
\nabla ^{2}\left( l_{2}^{2}\nabla \cdot \mathbf{B}_{c}-\left(
1-l_{2}^{2}\nabla ^{2}\right) \Psi \right) \equiv0
\end{equation*}%
and therefore the representation (\ref{G26}) is reduced to
\begin{equation}
\mathbf{u} =
\mathbf{B}_{c}+\nabla\times\mathbf\Psi
-\tfrac{1}{2}\nabla
\left(
\kappa \,\mathbf{r}\cdot \mathbf{B}_{c} 
+ \tfrac{l_{1}^{2}}{\mu }\mathbf{r}\cdot \mathbf{b}
- 2l_{1}^{2}\nabla \cdot \mathbf{B}_{c}+\left( \kappa
- l_{1}^{2}\nabla ^{2}\right) B_0
\right)
\label{G27}
\end{equation}

Let us apply the following transformations to the last term  in brackets:
\begin{equation}
\begin{aligned}
( \kappa- l_{1}^{2}\nabla ^{2}) B_0
&= (\kappa- l_{1}^{2}\nabla ^{2}) 
\left(B_c +\tfrac{l_{1}^{2}}{\mu }\mathbf{r}\cdot \mathbf{b}
+ B_g - \tfrac{l_{1}^{2}}{\mu }\mathbf{r}\cdot \mathbf{b}\right)\\
&= (\kappa- l_{1}^{2}\nabla ^{2}) 
\left(\phi_c 
+ \tfrac{2l_{1}^{4}}{\kappa}\nabla\cdot\textbf B_c
+ 2\alpha l_1^2\phi_g\right)\\
&= \kappa\,\phi_c 
- \tfrac{l_{1}^{2}}{\mu}\mathbf{r}\cdot \mathbf{b}
+ 2l_{1}^{2}\nabla\cdot\textbf B_c
-2 l_1^2\phi_g
\label{trans} 
\end{aligned}
\end{equation}
where we take into account decomposition for $B_0$ \eqref{G14}, definitions for $\phi_c$ \eqref{G18} and $\phi_g$ (\ref{G21}),   governing equations for the potentials \eqref{G12}$_1$, \eqref{G19}, \eqref{G22} and definitions for coefficients $\kappa$ and $\alpha$.

Substituting (\ref{trans}) into (\ref{G27}), we obtain
\begin{equation}
\mathbf{u} =
\mathbf{B}_{c}+\nabla\times\mathbf\Psi
-\tfrac{1}{2}\nabla
\left(
\kappa \,\mathbf{r}\cdot \mathbf{B}_{c} 
+\kappa\,\phi_c 
-2 l_1^2\phi_g
\right)
\label{res0}
\end{equation}

We also can replace the term $\nabla \times \mathbf{\Psi }$ by the vector potential $\mathbf{B}_{g}$ assuming that the divergence of this vector vanishes ($\nabla\cdot \mathbf{B}_{g}=0$) and that RHS in its governing equation \eqref{G12} should contain $\hat{\textbf b}$ instead of $\textbf b$. Consequently, previous relation \eqref{res0} takes the form:%
\begin{equation}
\mathbf{u}=\mathbf{B}_{c}+\mathbf{B}_{g}-\tfrac{\kappa}{2} \nabla \left(
\mathbf{r}\cdot \mathbf{B}_{c}+\varphi _{c}\right) +l_{1}^{2}\nabla \varphi
_{g},
\end{equation}%
which represents the generalized PN solution for the displacement field (\ref{pngs}).

\section{Completeness of general solutions}

The completeness of the general solution means it can represent any displacement field governed by equilibrium equations  \cite{gurtin1972,mindlin1936note}.
Following classical approach (see, e.g. \cite{gurtin1972, wang2008recent}), the completeness of general solutions in SGE can be proven considering relations between their potentials. The proof based on BG and PN solutions was presented in \cite{solyaev2024complete}. Let us extend this result to the all other kinds of SGE general solutions considered above.

We start with BG representation of general solution \eqref{bggs1}:
\begin{equation}
\mathbf{u}=\tfrac{1}{\alpha }\left( 1-l_{2}^{2}\nabla ^{2}\right) \nabla
 \nabla \cdot \mathbf{G} -\left( 1-l_{1}^{2}\nabla ^{2}\right)
\nabla \times \nabla \times \mathbf{G} 
\label{comp1}
\end{equation}
that can be equivalently rewritten as:
\begin{equation}
\bar \alpha\left( 1-\bar l_{1}^{2}\nabla ^{2}\right) \nabla
 \nabla \cdot \mathbf{G} -\left( 1-\bar l_{2}^{2}\nabla ^{2}\right)
\nabla \times \nabla \times \mathbf{G} = -\frac{\bar {\textbf b}}{\mu}
\label{comp2}
\end{equation}
where we introduce $\bar \alpha = 1/\alpha$, $\bar {\textbf b} = -\mu\textbf u$, $\bar l_1=l_2$ and $\bar l_2=l_1$.

Suppose further that displacement field $\textbf u$ (that persists in the right-hand side of equation \eqref{comp2}) is an arbitrary solution of the equilibrium equations \eqref{ee}.
Then we observe that equation \eqref{comp2} formally coincides with the form of equilibrium equations of SGE \eqref{ee} up to notation. Consequently, its solution can be represented by using any kind of general solution discussed above. In Ref. \cite{solyaev2024complete}, the generalized PN solution was used to define $\textbf G$ in terms of $\textbf u$. In the present case, we can employ the generalized TNH solution \eqref{I7}, \eqref{I13}, which is more compact:
\begin{equation}
\mathbf{G}=\bar{\textbf{P}} - \bar\kappa\nabla
\left(
    \mathcal N(\nabla\cdot\bar{\textbf P}) 
    - \tfrac{\bar l_2^2 - \bar l_1^2}{1-\bar\alpha} \mathcal H_1(\nabla\cdot\bar{\textbf P})
\right)
\label{comp3}
\end{equation}
where $\bar{\kappa} = (\bar\alpha-1)/\bar \alpha$ and the stress function $\bar{\mathbf P}$ is defined by:%
\begin{equation}
\begin{aligned}
\bar{\textbf P} = \mathcal N(\textbf u) 
+ \bar l_2^2 \mathcal H_2(\textbf u)
\end{aligned}
\end{equation}%
in which we take into account representation \eqref{P1}-\eqref{ghe} to define the particular solution of governing equation \eqref{I13}. 

Thus, we can derive an expression for the Galerkin vector function $\textbf G$ based on any predefined displacement field $\textbf u$ satisfying \eqref{ee}. This means that BG representation \eqref{bggs1} is complete within SGE. Given all the previously established relationships between the potentials of various general solutions in Section 4, we can assert that all these solutions are also complete. This includes the completeness of Mindlin's representation, which, to our knowledge, has not been previously discussed in the literature.

\section{Conclusion}
\label{con}

In this work, we have presented an overview and analysis of displacement general solutions for the equilibrium equations of SGE. We have shown that, in fact, all classical solutions can be generalized to SGE by extending the classical approaches used to construct these solutions. Furthermore, we demonstrated that essentially all classical solutions can be simply generalized to SGE by means of the universal representation \eqref{gd}-\eqref{gdeg}, which considers the simple sum of any classical general solution, the Helmholtz decomposition for gradient part of the displacement field, and additional term related to the potential of body forces.

In the introduction to this article, we provided a brief overview of known analytical and semi-analytical solutions that have employed displacement general solutions in SGE. We note that there are far fewer such examples in SGE compared to classical elasticity, which is due to the fact that their construction requires significantly more complex approaches, as well as the fact that many modern problems are solved by numerical methods. Nevertheless, the application of analytical general solutions can be extremely useful, such as in validating numerical methods based on closed-form solutions \cite{Shekarchizadeh2022}, deriving expressions convenient for processing experimental data \cite{khakalo2017gradient}, and for constructing and analyzing complicated asymptotic solutions \cite{solyaev2022elastic,solyaev2024higher}.

We note that this paper has yielded several new results, including the derivation of a general solution in the TNH form for SGE, the development of a general representation incorporating an arbitrary classical displacement field, and the proof of relationships between various general solutions. Namely, presented relationship between the BG and TNH solutions is non-trivial within SGE and has been obtained here for the first time. Of particular interest, in our view, is also the established equivalence between solutions of the Mindlin type and those of the generalized PN type (including LBVB solution). These two forms of the general solution have existed in parallel for two decades, but the connection between them has not been shown previously.

Currently, there is significant growth and development in the fields where SGE models find application -- in micromechanics \cite{gao2009green,delfani2025anti}, in problems of modeling metamaterials and composites \cite{forest2011generalized,dell2019pantographic,ganghoffer2025quasi}, in dislocation theory \cite{gutkin1999dislocations,lazar2006dislocations}, in fracture mechanics \cite{placidi2018energy,rao20263d}, in the mechanics of nanosized objects \cite{forest2020strain,coredo2016} etc. Both fundamental theoretical aspects, related to proving existence theorems and the correct formulation of boundary value problems \cite{eremeyev2018linear, lazar2015non,fedele2023kind}, and applied problems, related to the  identification of materials constants \cite{yuheng2024new, vasiliev2021new, rezaei2024procedure,solyaev2025evaluation,shodja2018toupin}, are being considered. 
The displacement general solutions presented in this work could be one of the convenient tools for addressing these challenges.

\section*{Appendix A. Derivations for generalized TNH solution}
\label{appa}

To obtain the final form of TNH solution (32)  in Section 3.6 the following steps should be performed. At first, we note that equation (29) can be obtained from (28) by using the following derivations:
\begin{equation}
\begin{aligned}
\mathbf{P}^{1} &=\mathcal{H}_{2}\nabla \mathcal{N}\left( \left(
1-l_{2}^{2}\nabla ^{2}\right) \nabla \cdot (\alpha \mathcal{H}_{2}\left(
1-l_{1}^{2}\nabla ^{2}\right) \mathbf{u}^{1}+\mathbf u^2)\right)   \\
&=\alpha\mathcal{H}_{2}\nabla \mathcal{N}\left( \left(
1-l_{2}^{2}\nabla ^{2}\right) \mathcal{H}_{2}  \left(
1-l_{1}^{2}\nabla ^{2}\right) \nabla \cdot\mathbf{u}^{1}\right)    \\
&=\alpha \mathcal{H}_{2}\nabla \mathcal{N}\left( \left( 1-l_{1}^{2}\nabla
^{2}\right) \nabla \cdot \mathbf{u}^{1}\right)    \\
&=\alpha \mathcal{H}_{2}\left( 1-l_{1}^{2}\nabla ^{2}\right) \nabla
\mathcal{N}\left( \nabla \cdot \mathbf{u}\right)   \\
&=\alpha \mathcal{H}_{2}\left( 1-l_{1}^{2}\nabla ^{2}\right) \mathbf{u}^{1}
\label{a1}
\end{aligned}
\end{equation}
where we take into account that $\mathbf{u}^{1} = \nabla
\mathcal{N}\left( \nabla \cdot \mathbf{u}\right)$ and $\nabla\cdot\textbf u^2 =0$ (see (22)). 

Then, from this result we find: 
\begin{equation}
\begin{aligned}
\mathbf{u}^{1} &=\tfrac{1}{\alpha}\mathcal{H}_{1}\left( 1-l_{2}^{2}\nabla
^{2}\right) \mathbf{P}^{1}
=\tfrac{1}{\alpha}\left(1-(l_{1}^{2}-l_{2}^{2}) \nabla ^{2}%
\mathcal{H}_{1} \right)\mathbf{P}^{1}
\label{a2}
\end{aligned}
\end{equation}

Substituting (30) and \eqref{a2} into (22) and using (27) we obtain:
\begin{equation}
\begin{aligned}
\mathbf{u} &=\mathbf{u}^1 +\mathbf{u}^2 =\tfrac{1}{\alpha }\left( 1-\left( l_{2}^{2}-l_{1}^{2}\right)
\nabla ^{2}\mathcal{H}_{1}\right) \mathbf{P}^{1}+\mathbf{P}^{2},\\[5pt]
&=\mathbf{P}+\left( \tfrac{1}{\alpha }\left( 1-\left(
l_{2}^{2}-l_{1}^{2}\right) \nabla ^{2}\mathcal{H}_{1}\right) -1\right)
\mathbf{P}^{1},   \\
&=\mathbf{P}-\kappa \nabla \left( 1-\tfrac{l_{2}^{2}-l_{1}^{2}}{1-\alpha }\nabla ^{2}\mathcal{H}_{1}\right) \mathbf{P}^{1} 
\label{a3}
\end{aligned}
\end{equation}

Taking into account that $\mathbf{P}^{1}  = \mathcal{N}( \nabla \cdot \mathbf{P})$ according to Helmhotlz decomposition, we obtain:
\begin{equation}
\begin{aligned}
\mathbf{u} 
&=\mathbf{P}-\kappa \nabla \left( 1-\tfrac{l_{2}^{2}-l_{1}^{2}}{1-\alpha }\nabla ^{2}\mathcal{H}_{1}\right) \mathcal{N}%
\left( \nabla \cdot \mathbf{P}\right)   \\
&=\mathbf{P}-\kappa \nabla\left( \mathcal{N}\left( \nabla \cdot \mathbf{P}\right) -\tfrac{l_{2}^{2}-l_{1}^{2}}{1-\alpha } \mathcal{H}_{1}\left( \nabla \cdot \mathbf{P}\right) 
\right)
\label{a4}
\end{aligned}
\end{equation}
that is the first variant of TNH solution (32).

The second variant of TNH solution can be obtained if we use the following form of equation \eqref{I4}:
\begin{equation}
\left( 1-l_{1}^{2}\nabla ^{2}\right) \nabla ^{2}\left( \alpha \mathbf{u}^{1}+
\mathcal{H}_{1}\left( 1-l_{2}^{2}\nabla ^{2}\right) \mathbf{u}^{2}\right) =-\frac{\mathbf{b}}{\mu}  
\label{I15}
\end{equation}

We next define the vector function within the brackets as the second variant of TNH potential:
\begin{equation}
\mathbf{P}=\alpha \mathbf{u}^{1}+\mathcal{H}_{1}\left( 1-l_{2}^{2}\nabla
^{2}\right) \mathbf{u}^{2}  
\label{I16}
\end{equation}%

Then from (\ref{I15}) and (\ref{I16}) it follows that
\begin{equation}
\left( 1-l_{1}^{2}\nabla ^{2}\right) \nabla ^{2}\mathbf{P}=-\frac{\mathbf{b}%
}{\mu }.  \label{I17}
\end{equation}

According to the Helmholtz decompositions (7), (8), we can introduce:
\begin{equation}
\mathbf{P}=\mathbf{P}^{1}+\mathbf{P}^{2},  
\qquad
\mathbf{P}^{1}=\nabla\mathcal{N}\left(\nabla \cdot \mathbf{P}\right),
\qquad
\nabla\cdot \mathbf{P}^{2} = 0
\label{I18}
\end{equation}%

Using (\ref{I16}) in (\ref{I18})$_2$ we find
\begin{equation}
\mathbf{P}^{1} 
=\alpha\nabla \mathcal{N}(\nabla \cdot \mathbf{u}^{1}) 
=\alpha \nabla \mathcal{N}\left( \nabla \cdot \mathbf{u}\right)   
=\alpha \mathbf{u}^{1}
\label{I20}
\end{equation}
where we take into account (22).

Thus, in the second variant of TNH solution, instead of a simple equality between the solenoidal parts of the displacement field and TNH potential ($\textbf u^2=\textbf P^2$ (30), which holds in the classical case as well), we have a simple proportionality for their irrotational parts \eqref{I20}.
Definition for the solenoidal part of $\textbf P$  can be obtained from (\ref{I16}), (\ref{I18}) and (\ref{I20}) in the following form:
\begin{equation*}
\mathbf{P}^{2}=\mathcal{H}_{1}\left( 1-l_{2}^{2}\nabla ^{2}\right) \mathbf{u}^{2}
\end{equation*}
which yields
\begin{equation}
\mathbf{u}^{2}=\mathcal{H}_{2}\left( 1-l_{1}^{2}\nabla ^{2}\right) \mathbf{P}%
^{2}  \label{I22}
\end{equation}%

Employing equations (22), (105), (106) and (107), we can obtain%
\begin{equation}
\begin{aligned}
\mathbf{u} &=\tfrac{1}{\alpha }\mathbf{P}^{1}+\mathcal{H}_{2}\left(
1-l_{1}^{2}\nabla ^{2}\right) \mathbf{P}^{2},   \\
&=\mathcal{H}_{2}\left( 1-l_{1}^{2}\nabla ^{2}\right) \mathbf{P}-\tfrac{1}{%
\alpha }\left( \alpha \mathcal{H}_{2}\left( 1-l_{1}^{2}\nabla ^{2}\right)
-1\right) \mathbf{P}^{1},   \\
&=\mathcal{H}_{2}\left( 1-l_{1}^{2}\nabla ^{2}\right) \mathbf{P}-\tfrac{1}{%
\alpha }\left( \alpha -1-\alpha \left( l_{1}^{2}-l_{2}^{2}\right) \nabla ^{2}%
\mathcal{H}_{2}\right) \mathbf{P}^{1},   \\
&=\mathcal{H}_{2}\left( 1-l_{1}^{2}\nabla ^{2}\right) \mathbf{P}-\kappa
\nabla \left( 1-\tfrac{l_{1}^{2}-l_{2}^{2}}{\kappa } 
\nabla^{2}\mathcal{H}_{2}\right) \mathcal{N}\left( \nabla \cdot \mathbf{P}\right) , \\
&=\mathcal{H}_{2}\left( 1-l_{1}^{2}\nabla ^{2}\right) \mathbf{P}-\kappa
\nabla \left(
\mathcal{N}\left( \nabla \cdot \mathbf{P}\right) 
-\tfrac{l_{1}^{2}-l_{2}^{2}}{\kappa }
\mathcal{H}_{2}\left( \nabla \cdot \mathbf{P}\right)
\right) 
\label{I23}
\end{aligned}
\end{equation}%
or equivalently%
\begin{equation}
\mathbf{u}=\left( 1-\left( l_{1}^{2}-l_{2}^{2}\right) \nabla ^{2}\mathcal{H}_{2}\right) \mathbf{P}
-\kappa
\nabla \left(
\mathcal{N}\left( \nabla \cdot \mathbf{P}\right) 
-\tfrac{l_{1}^{2}-l_{2}^{2}}{\kappa }
\mathcal{H}_{2}\left( \nabla \cdot \mathbf{P}\right)
\right)   
\label{I24}
\end{equation}

The representation of the displacement field (\ref{I23}) or (\ref{I24}) is the second variant of TNH general solution within SGE that was given in Section 3.6. The function $\mathbf{P}$ is the second kind of TNH potential that obeys the fourth-order governing equation (\ref{I17}).

\section*{References}
\renewcommand{\bibsection}{}
\bibliography{refs.bib}

\end{document}